\documentclass[preprint,5p,twocolumn,times,sort&compress]{elsarticle}
\pdfoutput=1                                                                                                                                                  

\newcommand{\CO}{subsurface CO$_2$}
\newcommand{\co}{CO$_2$}
\newcommand{\E}{$E$}
\newcommand{\dmax}{$d_{\text{max}}$}

\newcommand{\gcm}{g/cm$^3$}
\newcommand{\dpack}{$P_{\text{det}}$}
\newcommand{\bpack}{$N_{\text{bar}}$}
\newcommand{\bpackhit}{$N_{\text{bar}}^{\text{hit}}$}
\newcommand{\xreal}{$r_{\text{real}}$}
\newcommand{\xproj}{$r_{\text{proj}}$}

\newcommand{\mr}{\mathrm} 
\newcommand{\mb}{\mathbf}

\newcommand{\du}{\mathrm{d}}
\newcommand{\geant}{\textsc{Geant4}}

\usepackage{natbib}
\usepackage[normalem]{ulem}
\usepackage{multirow}
\usepackage{lineno}
\usepackage{todonotes}
\usepackage{graphicx}
\usepackage{caption}
\usepackage{subcaption}
\usepackage{dcolumn}
\usepackage[retainorgcmds]{IEEEtrantools}
\usepackage{sidecap} 
\usepackage{placeins}
\usepackage{amsmath}
\usepackage{cancel}
\usepackage{tabularx, booktabs}
\usepackage{gensymb}
\usepackage{upgreek}
\usepackage{nicefrac}

\newcommand*\patchAmsMathEnvironmentForLineno[1]{%
  \expandafter\let\csname old#1\expandafter\endcsname\csname #1\endcsname
  \expandafter\let\csname oldend#1\expandafter\endcsname\csname end#1\endcsname
  \renewenvironment{#1}%
     {\linenomath\csname old#1\endcsname}%
     {\csname oldend#1\endcsname\endlinenomath}}%
\newcommand*\patchBothAmsMathEnvironmentsForLineno[1]{%
  \patchAmsMathEnvironmentForLineno{#1}%
  \patchAmsMathEnvironmentForLineno{#1*}}%
\AtBeginDocument{%
\patchBothAmsMathEnvironmentsForLineno{equation}%
\patchBothAmsMathEnvironmentsForLineno{align}%
\patchBothAmsMathEnvironmentsForLineno{flalign}%
\patchBothAmsMathEnvironmentsForLineno{alignat}%
\patchBothAmsMathEnvironmentsForLineno{gather}%
\patchBothAmsMathEnvironmentsForLineno{multline}%
}

\newcolumntype{Y}{>{\centering\arraybackslash}X}

\title{Simulation of muon radiography for monitoring CO$_2$ stored in a geological reservoir}

\begin{document}


\author[sheffield]{J.~Klinger\corref{cor1}}
\ead{j.klinger@sheffield.ac.uk}
\cortext[cor1]{Corresponding author}

\author[durham_earth]{S.J.~Clark}
\author[nasa]{M.~Coleman}
\author[durham_energy]{J.G.~Gluyas}
\author[sheffield]{V.A.~Kudryavtsev}
\author[sheffield1]{D.L.~Lincoln}
\author[sheffield]{S.~Pal}
\author[boulby]{S.M.~Paling}
\author[sheffield]{N.J.C.~Spooner}
\author[sheffield]{S.~Telfer}
\author[sheffield]{L.F.~Thompson}
\author[sheffield]{D.~Woodward}

\address[sheffield]{Department of Physics and Astronomy, University of Sheffield, Sheffield, S3 7RH, UK}
\address[durham_earth]{Department of Earth Sciences, Durham University, Durham, DH1 3LE, UK}
\address[nasa]{NASA Jet Propulsion Laboratory, California Institute of Technology, Pasadena, California, USA}
\address[durham_energy]{Durham Energy Institute, Durham University, Durham, DH1 3LE, UK}
\address[sheffield1]{Department of Civil and Structural Engineering, University of Sheffield, Sheffield, S1 3JD, UK}
\address[boulby]{STFC Boulby Underground Science Facility, Boulby Mine, Cleveland, TS13 4UZ, UK}


\begin{abstract}
Current methods of monitoring \CO, such as repeat seismic surveys, are 
episodic and require highly skilled personnel to acquire the data.
Simulations based on simplified models have previously shown that 
muon radiography could be automated to continuously monitor \co~injection and migration, in addition to reducing the overall cost of monitoring.
In this paper, we present a simulation of the monitoring of \co~plume evolution in a geological reservoir using muon radiography.
The stratigraphy in the vicinity of a nominal test facility is modelled using geological data, and a numerical fluid flow model 
is used to describe the time evolution of the \co~plume. A planar detection region with a surface area of 1000~m$^2$ is considered, 
at a vertical depth of 776~m below the seabed.
We find that one year of constant \co~injection leads to
changes in the column density of $\lesssim 1\%$, and that the \co~plume is already resolvable with an exposure time of less than 50~days.
\end{abstract}

\maketitle

\section{Introduction}

The regulation of atmospheric greenhouse gas concentrations is required to stabilise the effects of anthropogenic climate change. 
Most scenarios of economic development predict a steep rise in greenhouse gas concentrations due to an increase in demand for energy, and hence
fossil fuel usage~\cite{emissions}. The option of \co~Capture and Storage (CCS) is regarded
as one technological solution 
for atmospheric \co~mitigation~\cite{ccs}.
CCS technologies have the capacity to reduce \co~emissions from power stations by up to 90\%~\cite{Stephenne20146106}, by the
injection of liquefied \co~into on-shore or off-shore geological repositories for permanent storage.

The principle of transporting and storing \co~in depleted oil fields or saline formations is already well understood, 
although there are technological and commercial issues which need to be addressed~\cite{Eiken20115541}.
In particular, there is a demand for a continuous and economically viable method for monitoring the injection and migration of \CO.
The migration of subsurface fluids can be highly unpredictable, even in developed oil fields~\cite{bailey}.
Current monitoring methods are episodic and require highly skilled personnel to acquire the data. 
Such technologies include repeat seismic surveys, measuring the subsurface response to electromagnetic waves, 
measuring specific gravity and monitoring pressure~\cite{techniques}. 
Any new, and preferably low-cost, technology 
which is able to provide automated and continuous monitoring will greatly enhance the long-term viability of CCS technologies.
One such option that has been proposed is muon radiography~\cite{Kudryavtsev201221}. 

High-energy muons, which are produced due to the interactions of cosmic-rays with the Earth's upper atmosphere, are deeply
penetrating charged particles. Muons are point-like charged particles which are very similar to electrons, but have a rest mass 
approximately 200~times that of the electron. 
All charged particles interact with other charged particles via the electromagnetic interaction. 
In this paper we consider muons interactions within matter, through which they are travelling. 
Although the materials in the Earth are electrically neutral,
the atoms within the materials are composed of positively charged nuclei and negatively charged electrons. When charged particles, such as cosmic-ray muons, 
interact with nuclei and electrons they scatter and lose energy by ionising and exciting atoms, 
as well as occasionally producing secondary particles that will be absorbed in matter quickly.
As with any stochastic energy transfer in a bulk material, the energy is eventually dispersed as heat.  

There are other cosmic-ray particles which one can consider at the surface of the Earth. Other point-like cosmic-ray particles such as electrons and photons\footnote{
Although photons are electrically neutral particles, they are the `force-carriers' of the electromagnetic interaction, and therefore by definition interact with electrically charged
particles.} lose energy 
at a much faster rate compared to muons underground, due having a significantly lower rest mass (or zero rest mass in the case of the photon) compared to muons.
Cosmic-ray particles which have a more complex internal structure (and are therefore not point-like) 
such as protons, neutrons, mesons and nuclei can additionally interact in the Earth via hadronic interactions. This 
additional interaction significantly reduces their range underground compared to muons. Therefore
muons are the only charged particles produced in the atmosphere which are observed in experiments based in underground observatories. There is a relatively large flux of neutrinos underground, although
their extremely low interaction probability means that they can be neglected.

Muons lose energy as they traverse dense materials and, once they have a sufficiently low energy,
will stop and be absorbed into the material or decay to produce an electron which will also be absorbed. 
Muon energy loss is fundamentally related to the density of the medium through which they
travel. It is also related to the mean atomic number and weight of the bulk material through which they pass. It is therefore possible to map the density profile of a large object by observing 
local variations in the flux of muons emerging from the target object. 

Muon radiography is already used for a wide variety of monitoring purposes, from relatively small-scale applications such as the
monitoring of cargo containers for high-$Z$ materials\footnote{The monitoring of high-$Z$ materials typically utilises the 
large scattering angle of muons incident on high mass nuclei, 
rather than changes in tshe muon flux due to muon attenuation, although Ref.~\cite{timblackwell} also uses muon disappearance.}~\cite{Borozdin1,Priedhorsky1,6728043,timblackwell}, 
to the mapping of density variations in large-scale structures such as voids in the ancient pyramids~\cite{alvarez1}
and magma chambers in volcanoes~\cite{Marteau201223,1995998}. Recent research has shown that muon radiography may be able to provide a low-cost, continuous 
\CO~monitoring system for the injection and subsequent movement of \co,
as an alternative or complementary system to seismic technology~\cite{Kudryavtsev201221}.

In this paper, we will show the results of the first detailed simulation
of muon radiography for this application. 
Our model incorporates geological data to describe a \CO~storage site, with a realistic model of \CO~plume evolution.
Such models will be an essential part of any future use of this technology. 
The experience of fluid injection processes in the oil industry 
suggests that natural systems do not necessarily behave as anticipated. An example of this is 
the use of waterflood to enhance oil production, which can lead to the 
rapid breakthrough of water without oil~\cite{bailey}. Thus, in 
the context of CCS, it will be essential to have a detailed model of expected behaviour
so that deviations from those expectations can be identified at the earliest stage.

It is clear that muon radiography will not be able to image \CO~with a higher resolution than technologies such as seismic surveying.
We propose, however, that muon radiography will be sufficiently accurate such that it may reduce the total cost associated with CCS, if used in combination with existing technology.
Therefore the utility of muon radiography as a tool to monitor \CO~will ultimately depend upon the practicality of deployment of the technology at \co~injection sites and 
the cost of deployment relative to other monitoring technology which could be used.  We will discuss these aspects in more detail in Section~\ref{disc}.

We will show that this technology is conceptually viable, and as such this paper 
represents a crucial milestone in the realisation of muon radiography for the purpose of CCS.

\begin{figure*}[t]
  \centering
  \includegraphics[width=0.75\linewidth]{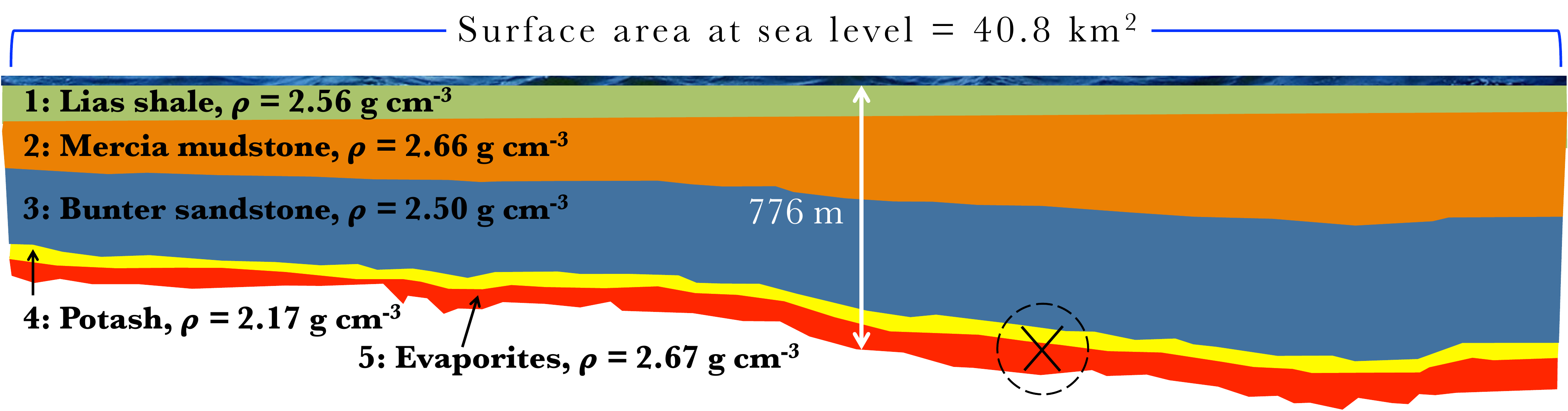}
  \caption{A diagram of geological stratigraphy model used 
  in this simulation. 
The layer number, simplified composition and average bulk density is indicated for each layer.
There is additionally a layer of sea water above layer 1, with a depth of 32~m. Layer `3' corresponds to the saline aquifer,
which is a suitable storage medium for \co. 
  The approximate location of the proposed detector site is indicated by the `X' symbol. In this simulation, \co~is piped underground via a vertical well, 
  with the point of injection vertically
   above the detector site.}
  \label{fig:geostrat}
\end{figure*}

\section{Conceptual overview}

Projected implementations of \co~capture technologies utilise pre-combustion, post-combustion~\cite{Desideri2010155} or 
oxyfuel processes~\cite{Mathieu2010283,Kluiters2010320}. Independent of the setup, liquefied or supercritical \co~can be subsequently
piped to an on-shore or off-shore geological storage site, of which there are already numerous examples~\cite{Eiken20115541}. In the context of this paper we consider off-shore sites, which may be more commercially attractive
due to the prospect of enhanced oil recovery~\cite{Blunt19931197,ehr}. 

In the current examples of this technology, liquefied or supercritical \co~is injected underground via a vertical well to a geological reservoir. 
It is required that the reservoir is sealed by an impermeable stratum (`cap-rock'),
in order to contain the \CO~volume, which will rise under buoyancy forces.

In this work, the geological storage reservoir considered is a saline aquifer, although the simulation procedures 
would be similar for a depleted oil or gas field with residual hydrocarbons and brine.
As a consequence of constant \co~injection, the density distribution of the reservoir will evolve. 
To model this, we use a numerical multiphase fluid flow model, which we describe in detail in Section~\ref{plume},
following the approach of Refs.~\cite{darren,Lewis1998a}. 
The model describes the vertical and horizontal distribution of \co~at discrete time steps, which evolves spatially
from the \co~injection point. Crucially, the model computes the change in the local density
due to \co~and brine distributions within the reservoir. As a result of the evolving density profile, which alters the 
survival probability of passing muons, one should expect 
localised modifications to the muon flux at depths below the storage reservoir. 

We propose that muon detectors are placed in containers that meet standard dimensions used for oil-well boreholes.
Full details of the detector dimensions are given in Section~\ref{mudet}. The detectors are assumed to be deployed in horizontal arrays below the reservoir, 
such that each array of detectors is fully contained within a surface area of approximately 1000~m$^2$. We assume for this study that 
each borehole container will contain a number 
of bars of plastic scintillator. Plastic scintillators, which emit light in response to the energy depositions of 
charged particles such as muons, are a low-cost and durable technology 
which is widely used in particle physics research.

Given the relatively low flux of muons at the depths
which are considered in this study, we employ a simple imaging strategy. We propose to resolve changes in the density of
the total geological section
via the change of the underground muon count due to \co~injection. Any background to the muon signals
underground can easily be suppressed by requiring a minimum number of scintillator bar hits. 
As the effect of \co~injection underground is to reduce
the overall density of the aquifer, one can measure an enhancement of the number of muons below the reservoir, with respect to the 
number of muons that one would expect in the absence of \CO.

\section{Geological modelling}
\label{geo}

An unfaulted geocellular model of the geology in the region of Boulby Mine, situated in Cleveland, United Kingdom,
has been produced for this study.  
There are a number of reasons why we have performed the simulation at this location, despite the site not being an 
option for off-shore CCS. Firstly, the geology of the site at Boulby is well understood. The stratigraphy
features a permian evaporite layer at a depth of 
0.75--1.1~km, which is typical of geological repository sites.
Furthermore, there is
safe, supported and versatile access provided 
by Israel Chemicals Ltd UK, who operate a commercial facility
in Boulby Mine,
and the STFC Boulby Underground Science 
Facility~\cite{doi:10.1080/10619127.2011.629920},
which allows for comprehensive testing of prototype detectors in the underground environment.

The geocellular model describes five strata below sea level, with a detector situated
776~m below the seabed, which is the approximate depth of a proposed subsurface muon detector
 test-site. The five strata are referred to 
sequentially as `layer 1', with the shallowest depth, to `layer 5', with the greatest depth. 
There is additionally a layer of seawater above layer 1, with a depth of 32~m vertically above the test-site.
The data sample which is used to describe the geological stratigraphy of this site is described in Section~\ref{test}, and the 
parameters which are used to describe the saline aquifer, as a hypothetical \co~storage site, are described
in Section~\ref{modelparam}.

Figure~\ref{fig:geostrat} depicts the model of the stratigraphy used in this simulation. The approximate location
of the detector site is indicated in the figure.
The implementation of the geological composition and strata boundary (`horizon') data in the simulation
framework is described in Section~\ref{geog4}. 

\subsection{Data sample}
\label{test}

Data describing four seismic horizons have been provided by Israel Chemicals Ltd UK. These horizons correspond to the top of layer~1 (the seabed), the top horizon of layer~4, the common 
horizon of layers~4 and~5, and the bottom horizon of layer 5. In order to represent the stratigraphy of layers 1--3, two 
additional layers were added to the model, based on the estimated thickness taken from well data. 

The horizon data are converted into a triangular-based mesh in order to reduce CPU and
memory overheads. The average distance between points on the mesh is approximately 100~m, which 
represents the level of accuracy of this model. The horizon data are combined with information describing the
composition of each layer. The average bulk density and simplified composition of each layer are indicated in Figure~\ref{fig:geostrat}. 
This information is 
used to populate each layer in the simulation described in Section~\ref{sim}. 

\subsection{Reservoir data}
\label{modelparam}

A numerical model 
is employed to describe the injection of \co~and the subsequent density evolution of the 
geological reservoir. 
The typical properties of the storage reservoir, which are required to parameterise the model, are described in this section.

The numerical model, which is described in detail in Section~\ref{plume}, is based on a system of equations 
which define a uniform axisymmetric region of the geological reservoir which fully
contains the volume of \CO. 
The fraction of the reservoir containing void spaces, initially occupied by the {\itshape in situ} brine, 
is given by the porosity $\phi$. 
The injected \co~is entirely confined above and below by impermeable horizontal boundaries, separated by a distance $H$. 
This also defines the height of the \co~injection well boundary, which is centred on the axis with a radius $r$. A constant mass-rate  of CO$_2$ injection, $M_\mr{n}$, is imposed which outwardly displaces the {\itshape in situ} brine. 

%
%
%

Accurate equations of state \cite{Bell2014,Rowe1970} are used with reference values of pressure and temperature for the reservoir system, in order to determine the corresponding fluid properties as listed in Table~\ref{t:APara}. The rock intrinsic density and porosity values are determined from a log of the test storage sandstone layer at the site of Boulby Mine, Cleveland, United Kingdom. The parameters which follow are typical experimental values characterising a sandstone-brine-CO$_2$ system following the study of Ref.~\cite{Mathias2013}, where further descriptions of their meaning are also given. Typical values for well depth below an impermeable boundary and well radius are given along with a feasible mass-rate of injection, which is at a lower limit of valuation deemed practical (3--120 kg/s) for commercial CCS purposes~\cite{Mathias2011a}. 

\begin{table}[t]
\renewcommand{\arraystretch}{1.2} 
	\footnotesize
    \centering
	\caption[]{The parameters used to model the \co~injection and multiphase interactions in the geological reservoir presented in this analysis. All parameters are assumed to have typical values, except for those indicated by `{\bf *}', which have been 
	acquired from data provided by Israel Chemicals Ltd UK.}\label{t:APara}
	\begin{tabular}{l l l l }
	\toprule
	Parameter                         		& Sym.             & Value    		& Units        \\ 
	\midrule
	Brine density                     		& $\rho_\mr{w}$    & 1100 & kg/m$^3$     \\
	Brine viscosity                   		& $\mu_\mr{w}$     & 900 & $\upmu$Pa\,s \\
	Brine bulk modulus                		& $K_\mr{w}$       & 2.90 & GPa          \\
	\midrule
	CO$_2$ density                    		& $\rho_\mr{n}$    & 720 & kg/m$^3$     \\
	CO$_2$ viscosity                  		& $\mu_\mr{n}$     & 60.0 & $\upmu$Pa\,s \\ 
	CO$_2$ bulk modulus               		& $K_\mr{n}$       & 0.025 & GPa          \\ 
	\midrule
	Reservoir rock intrinsic density ({\bf*})                      		& $\rho_\mr{s}$    & 2670 & kg/m$^3$     \\
	Porosity ({\bf*})                           		& $\phi$           & 0.15 & -            \\
	Permeability                      		& $\mb{k}$         & $1.875\times10^{-13}$	            & m$^2$        \\     
	Brine residual saturation         		& ${S\!}_\mr{rw}$      & 0.4438 & -            \\ 
	CO$_2$ end-point relative permeability  & $k_\mr{rn}$     & 0.3948 & -            \\ 
	Brine relative permeability exponent    & $m_\mr{k}$            & 3.2 & -            \\
	CO$_2$ relative permeability exponent   & $n_\mr{k}$            & 2.6 & -            \\
	van Genuchten parameter                 & $m_\mr{v}$  & 0.46 & -            \\
	van Genuchten parameter                 & $p_\mr{v}$       & 19.6 & kPa          \\
	\midrule
    Well/reservoir height             		& $H$              & 170 & m            \\
	Well radius                       		& $r$              & 0.2 & m            \\ 
	CO$_2$ mass-rate of injection        		& $M_\mr{n}$       & 20 & kg/s         \\
	\bottomrule
\end{tabular}
\end{table}


\section{Numerical \CO~simulation}
\label{plume}

Numerical models are widely used in order to solve large-scale fluid flow problems in porous media. 
In this section, we will present the formulation and implementation of a numerical model for the simulation of a geological 
storage reservoir over
discrete time steps after the start of a period of constant \co~injection. This model will then be used in Section~\ref{muo} to assess to what extent bulk density changes in such a system can be tracked over time using muon radiography.

We present a simple numerical model based on the coupled theoretical approach of Ref.~\cite{Lewis1998a}.
The model is then embedded into our \geant-based simulation\footnote{
\geant~is a simulation toolkit for modelling interactions and transport of elementary particles in matter, which
is used widely in the field of high energy physics.}~\cite{Agostinelli2003250}, described in Section~\ref{sim}, in order to account for the 
coupled interaction of \co~and brine within the reservoir,
 and hence the corresponding macroscopic changes in composition and density.

In Section~\ref{mathform}, we formulate the numerical model which we will use to generate \CO~plume formations,
using the input parameters described in Section~\ref{modelparam}. The bulk density distribution of the reservoir after \co~injection, 
which represent the solutions of this model, are discussed further in Section~\ref{dens_distribution}. 

One should note that we only intend for the model to be used in this simulation of muon radiography for the application of mapping density changes in a large overburden. It
is not intended that this model could be used for predictive purposes at this site. The site that we model is anyway unsuitable for \co~storage, rather it represents 
a site for which a complete dataset was available and, as discussed in Section~\ref{geo}, contains a serviced laboratory for testing prototype detectors.

\subsection{Mathematical formulation}
\label{mathform}

The storage of \co~in a deep saline aquifer, generally of moderate to high temperatures and pressures, presents a two-phase 
fluid system within the host rock; namely a `non-wetting' and `wetting' phase. The non-wetting phase,
modelled as a supercritical or liquid \co-rich fluid, displaces the wetting-phase, which is modelled as a liquid H$_2$O-rich fluid, within a porous solid (rock) phase. 

We assume that bulk composition and density changes are primarily due to the multiphase displacing and drainage behaviour within the reservoir 
system. We therefore concentrate on modelling the saturation effects, whilst neglecting the multiphase miscibility, 
thermal, inertial and solid deformation effects, which is typical for a first-order reservoir analysis.

Considering the continuity of multiple fluid phases in a porous medium,
 the following general macroscopic mass balance equation is given for each 
 phase$~\pi$,\footnote{In this formulation $\pi = \mr{n}, \mr{w}$, which subscript the terms belonging to the non-wetting and 
wetting phases respectively.}
\begin{equation}
\label{gen}
\frac{\partial \left( \phi S_\pi\rho_\pi \right)}{\partial t}
+\nabla\cdot\left( \phi S_\pi \rho_\pi \mb{v}_\pi \right) 
= 0,
\end{equation}
where $\phi$ is porosity (the fraction of void volume to total volume), $S_\mr{\pi}$ is fluid saturation 
(the fraction of fluid phase volume to void volume, such that $\sum_\pi S_\mr{\pi} = 1$), $\rho_\pi$ is intrinsic phase density and $\mb{v}_\pi$ is velocity.
The relevant constitutive relationships for bulk fluid compressibility and flow respectively are:
\begin{equation}
\label{con}
\frac{1}{\rho_\pi}\frac{\partial \rho_\pi}{\partial t}
=
\frac{1}{K_\pi}\frac{\partial p_\pi}{\partial t}
\quad
\text{and}
\quad
\phi S_\pi\mb{v}_\pi 
=
\frac{k_\mr{r\pi}\mb{k}}{\mu_\pi}\left(-\nabla p_\pi + \rho_\pi\mb{g}  \right),
\end{equation}

\noindent where $K_\mr{\pi}$ is bulk modulus, $p_\pi$ is fluid pressure, $\mb{g}$ is the gravity vector, and $\mb{k}$, $k_\mr{r\pi}$ and $\mu_\pi$ are the extended Darcy's law terms, for multiphase flow of intrinsic rock permeability, relative permeability and dynamic viscosity respectively. The constitutive relationships are substituted into Equation (\ref{gen}) by assuming $\phi$ constant and dividing through by $\rho_\pi$, whilst neglecting the gradient of  $\rho_\pi$, 
giving
\begin{equation}
\label{fld}
\frac{\phi S_\pi}{K_\pi}\frac{\partial p_\pi}{\partial t}
+ \phi \frac{\partial S_\pi }{\partial t}
+\nabla\cdot\left[ \frac{k_\mr{r\pi}\mb{k}}{\mu_\pi}\left(-\nabla p_\pi + \rho_\pi\mb{g}  \right) \right] 
= 0.
\end{equation}

The fluid saturation capacity relationship is introduced:
\begin{equation}
\label{sat}
\frac{\partial S_\mr{w}}{\partial t}
=
\frac{\partial S_\mr{w}}{\partial p_\mr{c}}
\frac{\partial p_\mr{c}}{\partial t}
=
\frac{\partial S_\mr{w}}{\partial p_\mr{c}}
\left(
\frac{\partial p_\mr{n}}{\partial t}
-
\frac{\partial p_\mr{w}}{\partial t}
\right),
\end{equation}
where $S_\mr{w}(p_\mr{c})$ is the saturation of the wetting phase, which is controlled by the capillary pressure $p_\mr{c}$. 
The capillary pressure is 
defined as the difference in pressure between the non-wetting and wetting phases, $p_\mr{c}=p_\mr{n}-p_\mr{w}$.

Employing (\ref{fld}) for both fluid phases gives two coupled equations on instating the appropriate fluid subscripts. Finally, substituting (\ref{sat}) into the coupled equations resolves the fluid pressures as the primary variables. In order to solve this system of equations, they are cast into the following spatially discretised form by employing the standard Galerkin finite element procedure \cite{Zienkiewicz2000}, within the domain $\Omega$ and on its boundary $\Gamma$, for which the usual initial and boundary conditions 
are incorporated~\cite{Lewis1998a}:
\begin{equation}
\medmuskip=0mu
\thickmuskip=0mu
\label{fe1}
\left[
\begin{array}{cc}
\mb{C}_\mr{w} & \mb{Q} \\
\mb{Q}^\mr{T} & \mb{C}_\mr{n} \\
\end{array}
\right]
\frac{\du}{\du t}
\left\{
\begin{array}{c}
\mb{p_\mr{w}} \\
\mb{p_\mr{n}}
\end{array}
\right\}
+
\left[
\begin{array}{cc}
\mb{H}_\mr{w}  & \mb{0}        \\
\mb{0}         & \mb{H}_\mr{n} \\
\end{array}
\right]
\left\{
\begin{array}{c}
\mb{p_\mr{w}} \\
\mb{p_\mr{n}}
\end{array}
\right\}
=
\left\{
\begin{array}{c}
\mb{f_\mr{w}} \\
\mb{f_\mr{n}}
\end{array}
\right\}
\end{equation}
where
\begin{align}
\label{fe2}
  \begin{aligned}
   \mb{C}_\pi &= \int_{\Omega} \mb{N}^\mr{T} \left(\frac{\phi S_\pi}{K_\pi}-\phi  \frac{\partial S_\mr{w}}{\partial p_\mr{c}}\right) \mb{N} \,\du\Omega, \\ \\
   \mb{Q}     &= \int_{\Omega} \mb{N}^\mr{T} \left( \phi \frac{\partial S_\mr{w}}{\partial p_\mr{c}} \right)                    \mb{N} \,\du\Omega, \\ \\
      \mb{H}_\pi &= \int_{\Omega} \left(\nabla\mb{N}\right)^\mr{T} \frac{k_\mr{r\pi}\mb{k}}{\mu_\pi} \nabla\mb{N} \,\du\Omega,\\ \\
   \mb{f}_\pi &= \int_{\Omega} \left(\nabla\mb{N}\right)^\mr{T} \frac{k_\mr{r\pi}\mb{k}}{\mu_\pi}\rho_\pi\mb{g} \,\du\Omega - \int_{\Gamma} \mb{N}^\mr{T} \frac{q_\pi}{\rho_\pi} \,\du\Gamma,
  \end{aligned}
\end{align}

\noindent are compressibility, coupling, permeability and supply matrices respectively. The $\mb{N}$ terms are a set of linear shape functions interpolating over the discretised domain. 
The mass flux, $q_\pi$, imposed normal to the boundary, is also introduced as a consequence of the boundary conditions.

To model the injection scenario, the system of equations (\ref{fe1}) are solved with the following specific boundary conditions. The inner (inflow) vertical injection boundary is prescribed with a mass-flux boundary condition, $q_\mr{n}$, which is related to the mass-rate of \co~injection by $M_\mr{n} = q_\mr{n}2\pi rH$. At the outer (outflow) radial extent of the domain, a constant hydrostatic pressure far-field boundary condition is prescribed. Otherwise, no-flux boundary conditions are assumed.

This system of equations is non-linear due to the dependence of the coefficient matrices on the primary variables; namely, the saturation and relative permeability terms, $S_\pi(p_\mr{c})$ and $k_\mr{r\pi}(S_\pi)$. These relationships are determined experimentally for a given system and are parametrised in this work by the van Genuchten $S_\pi$-$p_\mr{c}$ model~\cite{vanGenuchten1980} and by a power law $k_\mr{r\pi}$-$S_\pi$ relationship~\cite{Mathias2013}, respectively (see Table~\ref{t:APara}). The model parameters, $\phi$, $\mathbf{k}$, $\rho_{\pi}$,
$\mu_{\pi}$ and $K_{\pi}$ are assumed constant and uniform, which is typical for basic hydro-geological investigations.

Once the primary pressure variables are determined for a given point in time, the corresponding secondary saturation variables $S_\pi(p_\mr{c})$ are determined. From this, the following macroscopic bulk density can be given by summing the intrinsic phase densities multiplied by their corresponding volume fractions,
\begin{equation}
\label{e:VolFrac}
\rho_\mr{b} = (1-\phi)\rho_\mr{s} + \phi S_\mr{n}\rho_\mr{n} + \phi S_\mr{w}\rho_\mr{w},
\end{equation}
which introduces the intrinsic density of the solid rock/minerals, $\rho_\mr{s}$.

The spatial integrations of (\ref{fe2}) are carried out using Gaussian quadrature. The temporal integration of (\ref{fe1}) is carried out using finite differencing; the set of equations are solved monolithically using a fully implicit (unconditionally stable) embedded scheme producing solutions of adjacent order in accuracy to allow for error control and adaptive time-stepping. This is 
considered appropriate given that a coupled injection scenario is being modelled. 
Each time-step is solved iteratively due to the non-linearities, which is done via an accelerated fixed-point type procedure until convergence is met. 
The linearised system is solved at each iteration by a standard direct multi-frontal solver. 
Further details on the implementation of a system of equations of this type
are given in Refs.~\cite{Bathe1995,Lewis1998a,Zienkiewicz2000}.

\subsection{Reservoir bulk density distribution}
\label{dens_distribution}

The solutions to the model presented in Section~\ref{mathform} describe the 
density distribution of the reservoir at time~$t$ after the start of continuous \co~injection. 
The radial extent $r_\mr{E}(t)$ of the 
expanding volume of CO$_2$ has the relationship $r_\mr{E}\propto\sqrt{t}$~\cite{Nordbotten2005}.

It is found that density changes of
$\lesssim 4\%$ are expected due to the presence of injected CO$_2$, with respect to the bulk density of the reservoir pre-injection. For the purposes of muon
radiography, which is sensitive to the density of the total geological section, this corresponds to a change in the total column
density of $\lesssim 1\%$. In Section~\ref{sim}, Figure~\ref{fig:vox} shows a graphical representation of the subsurface bulk density
distribution 
at a specific point in time. 
This distribution is subsequently interfaced with \geant, following the voxelisation procedure presented in Section~\ref{s:vox}.


\section{Simulation framework}
\label{sim}

The simulation framework that is used for 
muon transport, in which muons descend through the geological model from Section~\ref{geo}, is described in this section.
All stages of the simulation are performed within the \geant.9.6 framework~\cite{Agostinelli2003250}.
In Section~\ref{geog4}, we present the strategy that we employ for converting the geological data to
\geant~volumes. In Section~\ref{s:vox}, we describe the implementation of the density distribution, $\rho_\mr{b}$, of the
reservoir due to the presence of
\co. In Section~\ref{muo}, we present the modelling of the muon flux distribution that we use at sea level, and the subsequent transport to the detector site.

\subsection{Geological stratigraphy interfacing}
\label{geog4}

The model of the stratigraphy of the test-site presented in Section~\ref{geo}
is interfaced with \geant. The implementation of the geological model in \geant~is highly CPU and memory intensive, so steps are taken to reduce such overheads.

In the simulation, each stratum of rock is constructed independently from one another within \geant,
which allows for multi-threading to be used in order to
reduce total computation time. We construct each stratum of rock using two parallel (and very approximately horizontal) meshes, 
bound by four vertical faces. Each mesh represents one of the horizons described in Section~\ref{geo}. 
We opt to construct the less detailed horizons using quadrangular-based meshes and the highly 
detailed horizons using triangular-based meshes.

The muon transport simulation, which is described in detail in Section~\ref{muo},
is performed in two stages: firstly only transporting muons to the lower horizon of layer 2 (or, equivalently, the top horizon of layer 3); 
and in a subsequent stage of simulation, transporting muons through the remaining three layers. One reason for this, in addition 
to a general improvement in CPU performance, is that fewer constructed strata are used in \geant, which
therefore reduces the memory overhead. Furthermore, as the first stage of the simulation only considers muons 
above the \CO~formation, it is sufficient to perform this stage of the simulation only once and then perform the second stage of the
simulation once for each time step of \CO~migration.

\subsection{Subsurface bulk density voxelisation}
\label{s:vox}

The numerical model describing the variation in bulk density of the reservoir, 
presented in Section~\ref{plume}, is implemented in the simulation 
using the voxelisation capabilities of \geant. In this framework, a regular grid of voxels (each of size $\{10\times10\times10\}$~m$^3$) populates the domain of interest, 
which is rendered and parameterised by the numerical model. 

We take steps to preserve the accuracy of the numerical model in the voxelisation procedure, particularly in regions of 
high numerical gradients caused by the fluid interface, which are due to the displacing behaviour of the \co.
The first stage of the procedure is to overlay the finite element mesh, which describes density profile and gradient with high precision, 
onto the voxel grid. The voxels are then parameterised with the composition and density outputs, which are 
interpolated from the finite element mesh at all of the overlapping voxel centroids. 

The voxelisation process is carried out at selected points in time. As described in Section~\ref{dens_distribution} and
Ref.~\cite{Nordbotten2005},
the radial extent $r_\mr{E}(t)$ of the \co~distribution expands with time $t$ as $r_\mr{E}\propto\sqrt{t}$. 
Accordingly, the voxelisation is performed at intervals $(2n)^2$ days which are therefore 
assumed to represent the reservoir state for the period between $(2n-1)^2$ and $(2n+1)^2$ days, where $n = 1\rightarrow 10$. 
For this simulation, using the parameters presented in Table~\ref{t:APara}, the radial extent is 
increased by approximately 36~m at each time-step until reaching a distance of approximately 360~m from the well. 

\begin{figure}[t]
  \centering
\includegraphics[width=\linewidth]{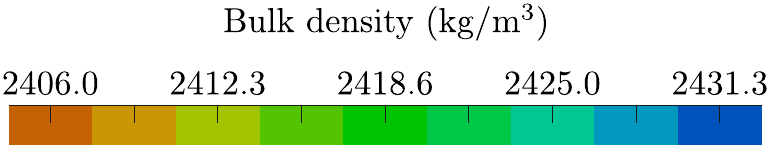}
\begin{subfigure}[b]{0.48\linewidth}
\includegraphics[width=\linewidth]{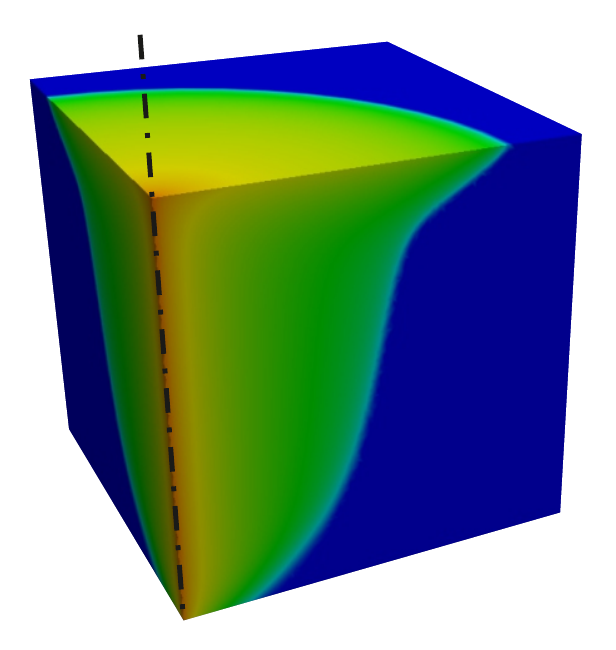}
\caption{\footnotesize Before voxelisation}
\end{subfigure}
\begin{subfigure}[b]{0.48\linewidth}
\includegraphics[width=\linewidth]{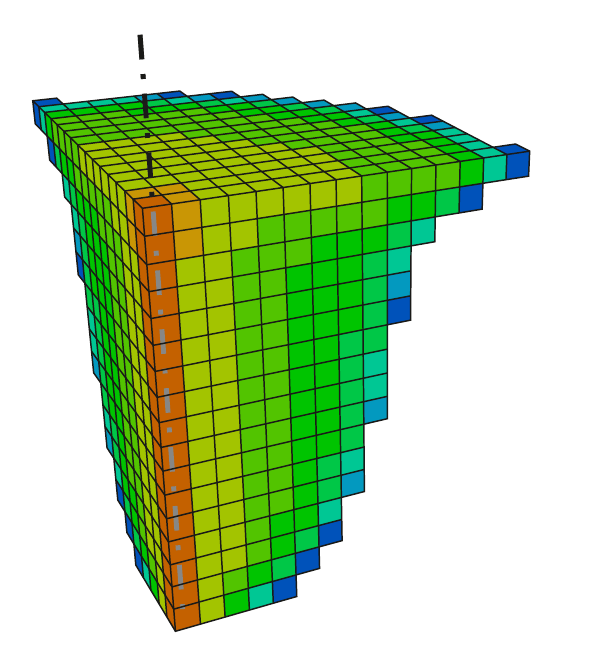}
\caption{\footnotesize After voxelisation}
\end{subfigure}
  \caption{Quarter-symmetric rendering of the sandstone-brine-CO$_2$ system bulk density as predicted after 64 days (a), alongside its equivalent voxelisation
of density bins for implementation with \geant~(b).}
  \label{fig:vox}
\end{figure}

A quarter-symmetric rendering of the system bulk density as predicted at 64 days, alongside its equivalent voxelisation
of density bins for implementation with \geant, is shown in Figure~\ref{fig:vox}.

\subsection{Muon flux sampling and particle transport}
\label{muo}

For the simulation of muon energy loss in solid material, 
we use the \geant.9.6 `shielding' physics list~\cite{Agostinelli2003250}, with the muon-nuclear interaction process explicitly included. 

We have implemented a Monte Carlo generator to sample
the spectrum of muon energies $E$ at sea level, as a function of the angle $\theta$ 
of the muon trajectory from the zenith.
The spectrum is based on 
the Gaisser parameterisation~\cite{gaisser1990cosmic}, which  
accounts for correlations between the $\theta$ 
and $E$ due to different muon-production mechanisms in the atmosphere. 
The parameterisation accounts for muon energy losses in the atmosphere, which are very low compared to the energy losses due to interactions underground.
As we describe later, we only consider muons with $E > 400$~GeV so we do not include recent measurements of the muon spectrum such as Ref.~\cite{bonechi} in 
which the spectrum is only measured up to 100~GeV. Muons which have been generated according the Gaisser parameterisation and subsequently
propagated to these depths have been previously shown to agree with 
data within $< 10\%$~\cite{Kudryavtsev2009339}.
We apply additional corrections to the parameterisation to account
for the muon lifetime and the Earth's curvature. 

In order to reduce computation time, a number of optimisations are implemented. 
There are no charged particles, other than muons, which can survive through large depths of matter. Therefore 
any secondary particles\footnote{
We define `primary' particles as those which are present at sea level,
and `secondary' particles as those which are produced at any depth below this.
} 
that are produced due to the interactions of muons with matter are immediately removed
from the simulation.

\begin{figure}[t]
  \centering
  \includegraphics[width=\linewidth]{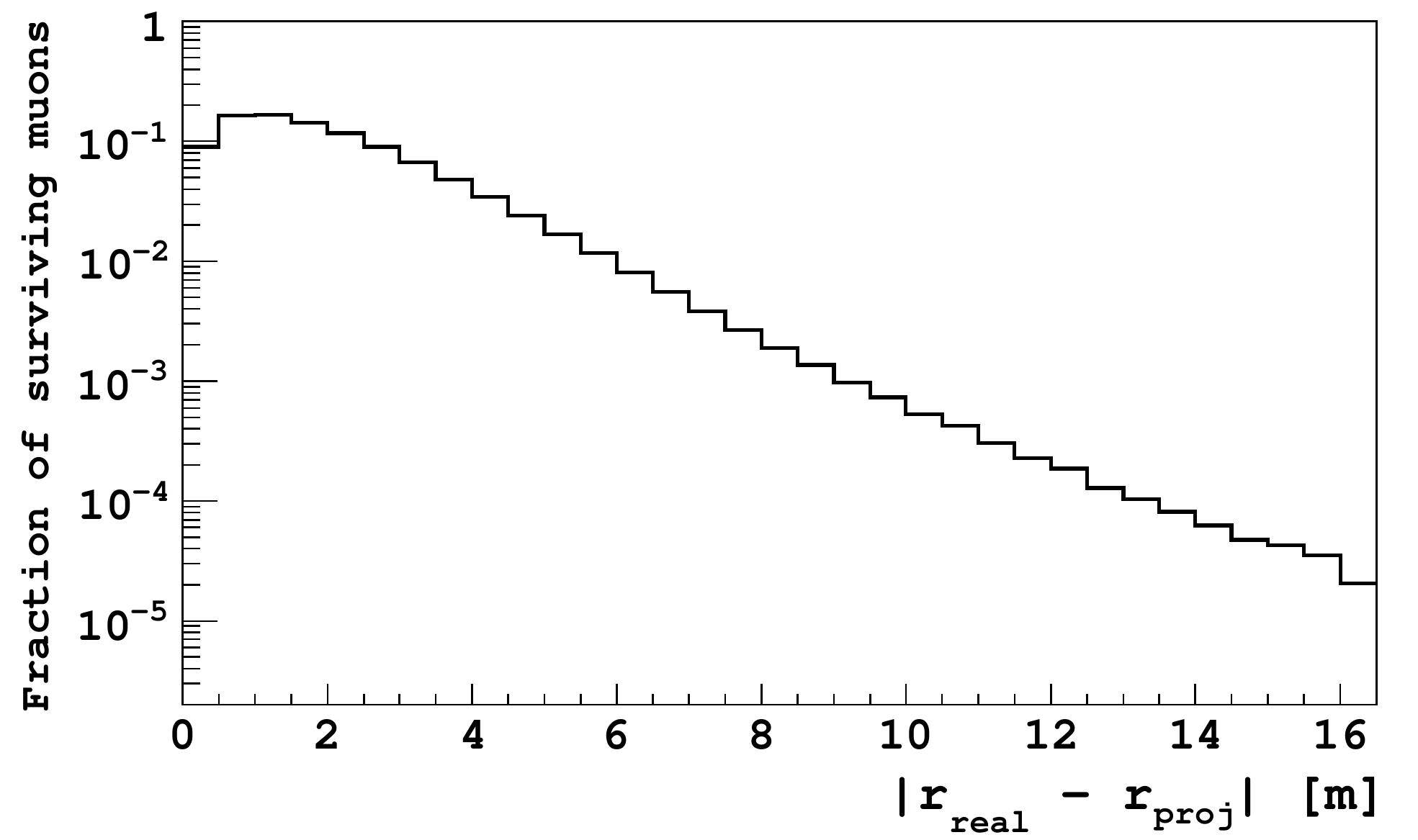}
  \caption{The distribution of $|$\xreal$-$\xproj$|$ between the horizontal muon position at the detector site, \xreal, and the position that is linearly projected from the muon's trajectory at sea level, \xproj.}
  \label{fig:displacement}
\end{figure}

We only consider muons with $E > 400$~GeV and $\theta < 70 \degree$, based on a previous
study of the muon survival depth~\cite{Kudryavtsev2009339,kud2}. 
Given the detector depth of $\sim 800$~m below sea level, and that the detector array is confined to within an area of 1000~m$^2$,
we require that muons originate from an approximately square region of $\{4.5 \times 4.5\}$~km$^2$ at sea-level.
This surface area is sufficiently large to account for any displacement of the muon's measured
position \xreal~due to scattering during the muon transport,
 with respect to the position that is linearly projected from the muon's trajectory at sea level, \xproj. 
 The distribution of $|$\xreal$-$\xproj$|$ of muons in this analysis are shown in Figure~\ref{fig:displacement}.
 We find that $|$\xreal$-$\xproj$|$ is a steeply falling distribution, with the average displacement of muons being $< 2.1$~m.

As a simulated muon loses energy, we re-evaluate the muon's maximum survival depth, \dmax, using a look-up table.
The look-up table, which maps \E~to \dmax, is generated 
 with the {\sc MUSIC} muon simulation code~\cite{Kudryavtsev2009339,kud2}. We then redefine \dmax, as a function of $E$, as the distance
 beyond which a muon has a survival probability of less than $10^{-6}$. The {\sc MUSIC} code only 
 considers materials with a uniform density, so 
 we conservatively choose to evaluate \dmax~for a material with density $\rho = 2.17$~\gcm, which 
 corresponds to the lowest density of the five strata (layer 4).
 We therefore remove muons at any point in the simulation which have a remaining transport distance greater than \dmax.
 One should note that the uniform density assumed in the MUSIC simulation code is only used to provide a conservative optimisation to the simulation 
 and does not affect the detail of our model.

\begin{figure}[t]
  \centering
  \includegraphics[width=\linewidth]{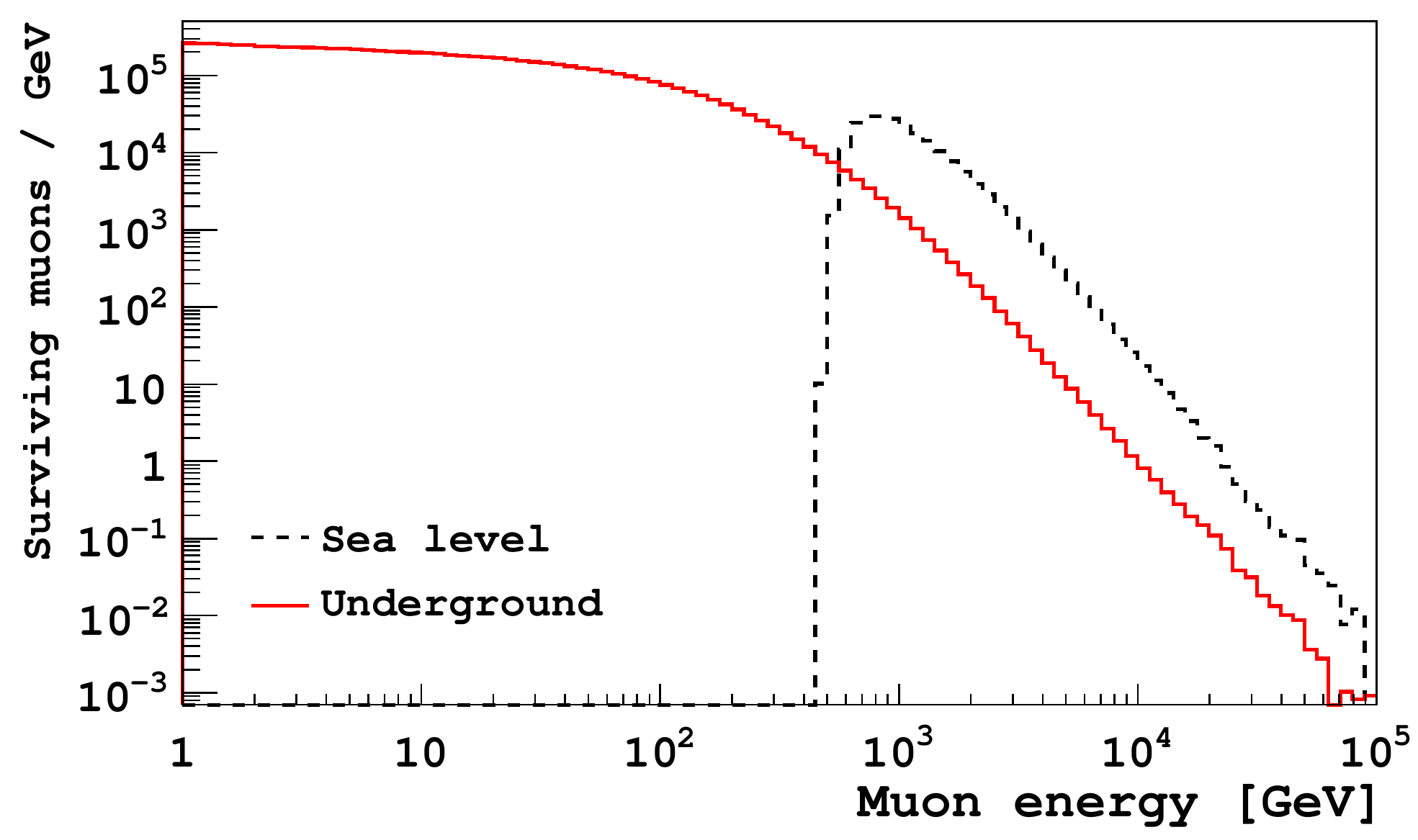}
  \caption{The energy spectra of muons, at sea level and underground respectively, for muons which
  survive to the detector. In total, approximately $140$~days of simulation data is presented, which corresponds to approximately 2.8$\times10^{7}$ muons in each spectrum.
  The same muons are used in both distributions, therefore only muons at sea level with energy greater than 400~GeV are observed underground. 
}
  \label{fig:muonspec}
\end{figure}

The energy spectra of muons, at sea level and underground respectively, for muons which
survive to the detector are shown in Figure~\ref{fig:muonspec}. The average energy of muons 
which reach the detector site is found to be approximately 1500~GeV at sea level, and 220~GeV underground.
The flux of muons at the detector site is found to be approximately $2.3 \times 10^{-7}$cm$^{-2}$s$^{-1}$.

\section{Muon detectors}
\label{mudet}

In this section, we discuss the general configuration of muon detectors that we propose for mapping 
changes in the density of the total geological section. In the analysis of the muon simulation data,
we present the maximum possible signal; therefore neglecting gaps between the detectors and assuming 
100\% acceptance for detecting muons whilst rejecting background from radioactivity and detector noise. The discussion
presented in this section is therefore intended to provide a degree of quantification of the total efficiency 
of these effects, whilst still allowing us to present our results in a manner that is independent of the detector configuration.

\subsection{Detector parameterisation}

In order to ensure the long-term stability of the monitoring system, we assume that plastic scintillators, coupled to 
appropriate photo-sensors, will be employed as muon detectors for this application.
We assume that individual detectors will be confined to containers capable of being deployed in
standard oil-well boreholes. 
The borehole containers would typically have an inner diameter of 15~cm and sufficient length (less than a few metres) for
commercially available plastic scintillator bars. Plastic scintillator bars can be acquired commercially with 
a cross-sectional area of $\{2 \times 2\}$~cm$^2$ and a length of typically $100$~cm. It is envisaged that multiple scintillator bars
can be packed into borehole containers. 
The borehole containers can be deployed in a horizontal array at some depth below the reservoir.

Plastic scintillators emit light in response 
to charged particles, such as muons. The light signal is converted into an electrical signal with photo-sensors, 
such as photomultipliers, and can be subsequently
digitised for analysis. 
The position of a muon `hit' along a scintillator bar can be inferred from the time difference between
signals at either end of the bar. 
A linear fit to multiple muon hits can be used to reconstruct a muon track. The relative position of each of the bars
gives the muon angle in the plane perpendicular to the long sides of the bars. The muon's angle in 
the plane parallel to the long sides of the bars can be calculated using the relative positions of the respective hits along each bar,
inferred from the timing information.

From geometric arguments, the angular resolution in the plane of circular borehole faces is approximately $3\degree$ for five scintillator bar hits. 
Position resolution based on timing and amplitude information from optical sensors leads to an angular resolution down to $5\degree$ in the plane 
along the long side of the bars. We consider angular cells of $7\degree \times 36\degree$ in Section~\ref{ana}.

For this study we assume 100\% trigger efficiency. 
The results that we present in Section~\ref{ana} may be scaled accordingly, once values of trigger efficiency 
are acquired for a particular detector configuration. 
We consider a single plane of muon detectors confined within a surface area of approximately 1000~m$^2$.
The limiting factor for this choice of surface area is the computation time associated with muon transport.

\subsection{Detector efficiency}
\label{deteff}

As there will be multiple scintillator bars per borehole container, 
a muon traversing a borehole container will give rise 
to multiple bar hits. Background signals, for example due to photo-sensor noise or radioactivity in the surrounding rock, 
which are assumed to be short-ranged or highly localised, can be suppressed by requiring a minimum number of bar hits \bpackhit.

The efficiency for muons to satisfy the condition on \bpackhit~is clearly dependent on the number and configuration of
scintillator bars in the borehole containers. Furthermore, the total acceptance $A$ for all muons in the detector region
must also account for
the fraction of the total surface area
in the 1000~m$^{2}$ detector region that actually contains borehole containers.
We therefore calculate the detector acceptance by considering two configuration parameters; \dpack, the fractional surface area occupied by detectors contained within the 1000~m$^{2}$ region, and \bpack, the number of scintillator bars packed into each borehole container. 
We consider two values of each parameter which, for both parameters, we refer to as `loose' and `tight':
\begin{itemize}
\item \dpack~$ = \{\text{loose} \sim 50\%, \text{tight} \sim75\%\}$,
\item \bpack~$ = \{\text{loose} : 16,     \text{tight} : 24\}$.
\end{itemize} 
It is assumed that the detectors are arranged homogeneously throughout the detector region. In reality, as discussed in Section~\ref{disc},
the muon detectors will be deployed in a number of boreholes (approximately 18), which extend radially from a mother borehole. 
Practically speaking therefore, the detectors will not be arranged in a square arrangement of approximately $\{30\times30\} \approx 1000$~m$^2$, as in our simulation. 
Instead we envisage the total instrumented area may occupy up to 1000~m$^2$. The choices of \dpack~are simply 
chosen to show the effect of the reducing/scaling the total number of observed muons from the maximum anticipated yield.

\begin{figure}[t]
  \centering
\begin{subfigure}[b]{0.48\linewidth}
\includegraphics[width=\linewidth,angle=90]{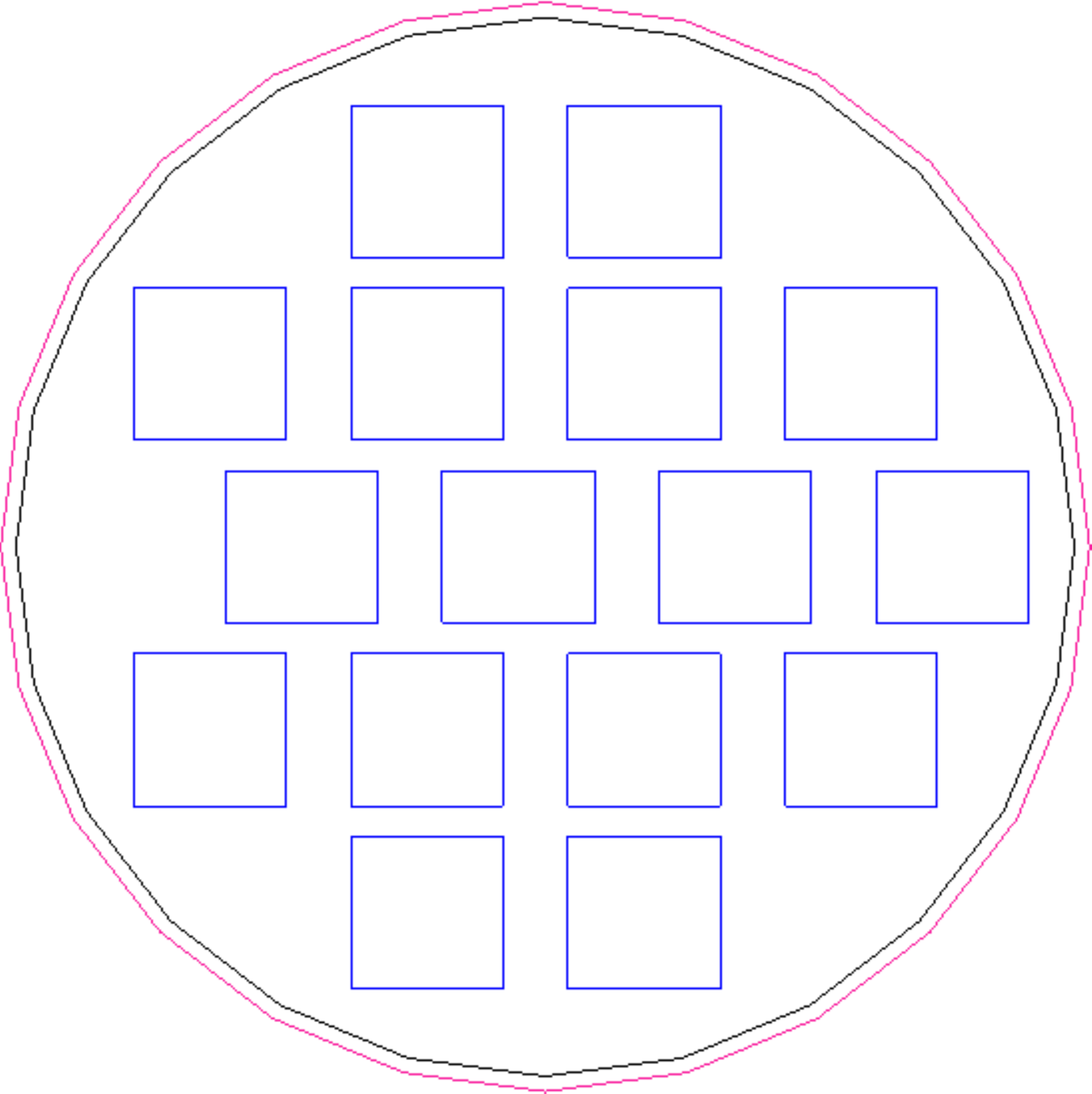}
\caption{\footnotesize Loose bar packing (16 bars)}
\end{subfigure}
\begin{subfigure}[b]{0.48\linewidth}
\includegraphics[width=\linewidth,angle=90]{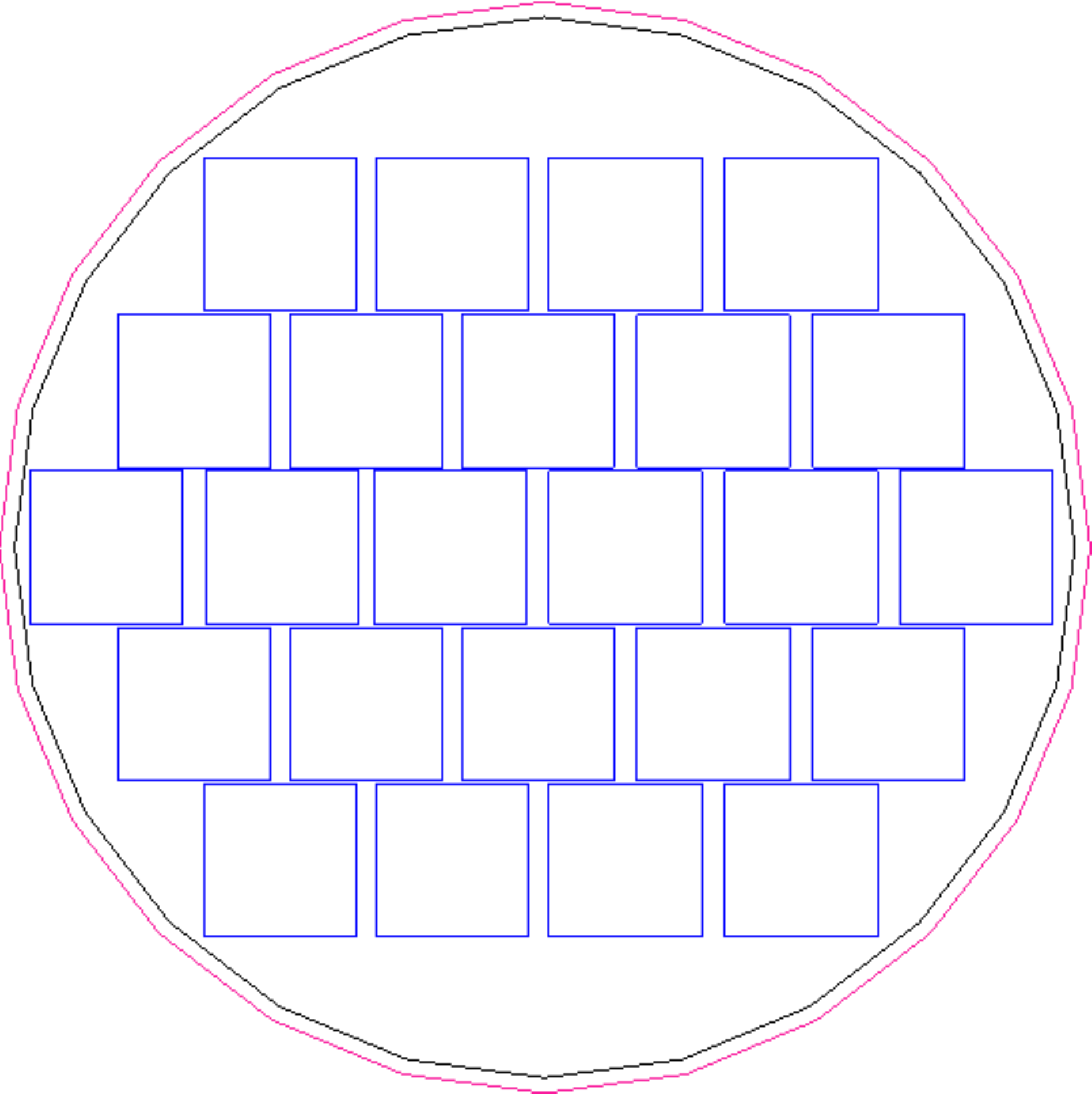}
\caption{\footnotesize Tight bar packing (24 bars)}
\end{subfigure}
  \caption{The arrangement of bars that we consider in this simulation, for the `loose' (a) and `tight' (b) values of \bpack.}
  \label{fig:dets}
\end{figure}

\begin{figure}[t]
  \centering
  \includegraphics[width=\linewidth]{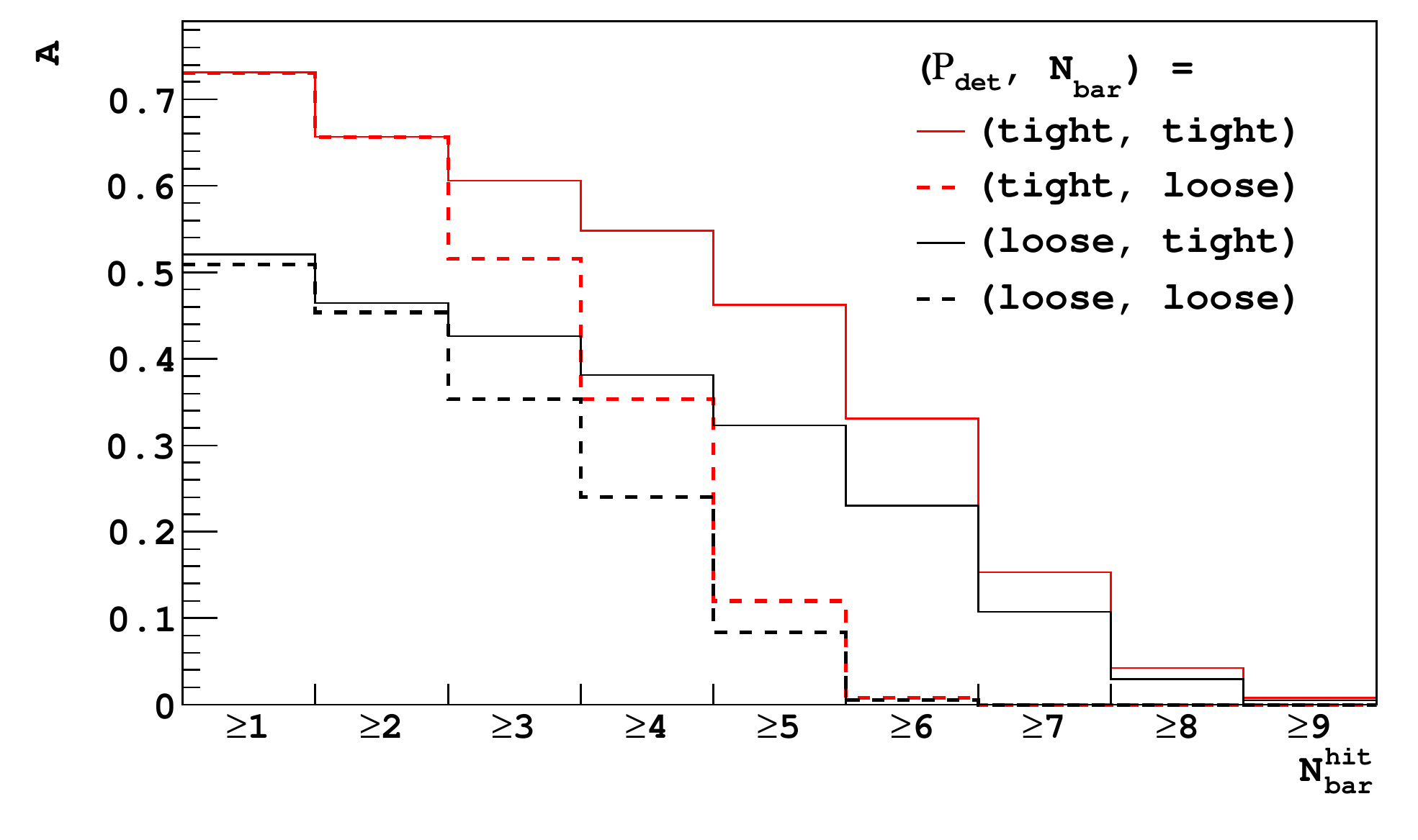}
  \caption{The acceptance of muons which
are recorded in the detector region in one plane of detectors, for all four combinations of \dpack~and \bpack.
The acceptance is shown as a function of the number of bars traversed by a muon,~\bpackhit. 
}
  \label{fig:acceptance}
\end{figure}

The arrangements 
of scintillator bars in the 
radial plane of the borehole containers,
that we consider for the `loose' and `tight' values of \bpack, are shown in Figure~\ref{fig:dets}. The acceptance of muons for a single
plane of detectors, $A$, which
are recorded in the detector region is shown in Figure~\ref{fig:acceptance}, for all four combinations of \dpack~and \bpack.
The acceptance $A$ is shown as a function of the \bpackhit, for surviving muons at the detector site crossing 
a single plane of detectors. Realistically, $\geq 3$ bars may be required to be hit to remove radioactive background and photo-sensor noise.

\section{Analysis of muon flux distribution}
\label{ana}

In this section, an analysis of the muon simulation data is performed, using 
muons which have been transported through the geological 
strata in the \geant~framework, as discussed in Section~\ref{sim}.

We propose a straightforward analysis strategy, in which we infer changes in the density of the total geological section
simply from the change in the number of detected
muons, with respect to the expectation in the absence of \CO. 
It is assumed that background noise and signals will be negligible once a condition is chosen for 
\bpackhit.\footnote{There is no known effect from electrical noise, radioactivity, or other sources that can lead to an appreciable background over muons at these depths, once the requirement for \bpackhit~is introduced.}
We consider increasingly large time steps in which muon data is collected, 
in sequential time periods since first \co~injection, as described in Section~\ref{s:vox}. 
This utilises the relationship between the radial extent of the \co~plume ($r_\mr{E}\propto\sqrt{t}$), such that the time intervals
correspond to periods under which the radius of the \co~plume is approximately constant.
We parameterise the significance $S$ of the change in the number of muons observed in a given time interval as:
\begin{equation}
S = \frac{N^{\text{after}} - N^{\text{before}} }{\sqrt{N^{\text{after}} + N^{\text{before}} }} \label{eq:significance}
\end{equation}

\noindent where $N^{\text{before}}$ and $N^{\text{after}}$ are the number of muon events observed over equal lengths of time, 
before any \CO~is injected and after \CO~is injected respectively. 
The number $N^{\text{before}}$ is calculated using a statistically independent sample of muons from those used to calculate $N^{\text{after}}$.
It is assumed that systematic uncertainties cancel in Equation~\ref{eq:significance}. The denominator in Equation~\ref{eq:significance}
is equal to the statistical uncertainty of the measurements.

In this analysis, both $N^{\text{before}}$
and $N^{\text{after}}$ refer to the number of muons detected at any point in the 1000~m$^2$ detector region.
In reality, the number of muons entering this calculation is reduced primarily due to the geometric factors discussed in 
Section~\ref{mudet}.
The effect of the true values of the \dpack~and \bpack~is 
to scale the observables $N^{\text{before}}$ and 
$N^{\text{after}}$ by an efficiency, $A$. 
It is therefore clear that $S$ will be reduced by a factor of $\sqrt{A}$. We do not apply factors of $\sqrt{A}$ to the values of $S$ presented in this section, as realistic values of \dpack~and \bpack~need to be investigated in further study. 
A degree of quantification of the acceptance $A$ can be inferred from
Figure~\ref{fig:acceptance}. 

The muon flux $\Phi_\mu$ at each time step $T$, and the associated significance $S$ (measured in standard deviations) 
of the change in muon count are shown in Table~\ref{tab:plume_tab}. 
The presented uncertainties are taken from the statistical uncertainty in the muon count.
A change in the global muon flux after 49 days corresponds to approximately 24 Gaussian standard deviations. This is clearly significant even for an acceptance of $A=5\%$.

\begin{figure}[h]
    \begin{subfigure}[b]{0.75\linewidth}
        \centering
        \begin{subfigure}[b]{0.47\linewidth}
            \centering        
            \includegraphics[width=\linewidth]{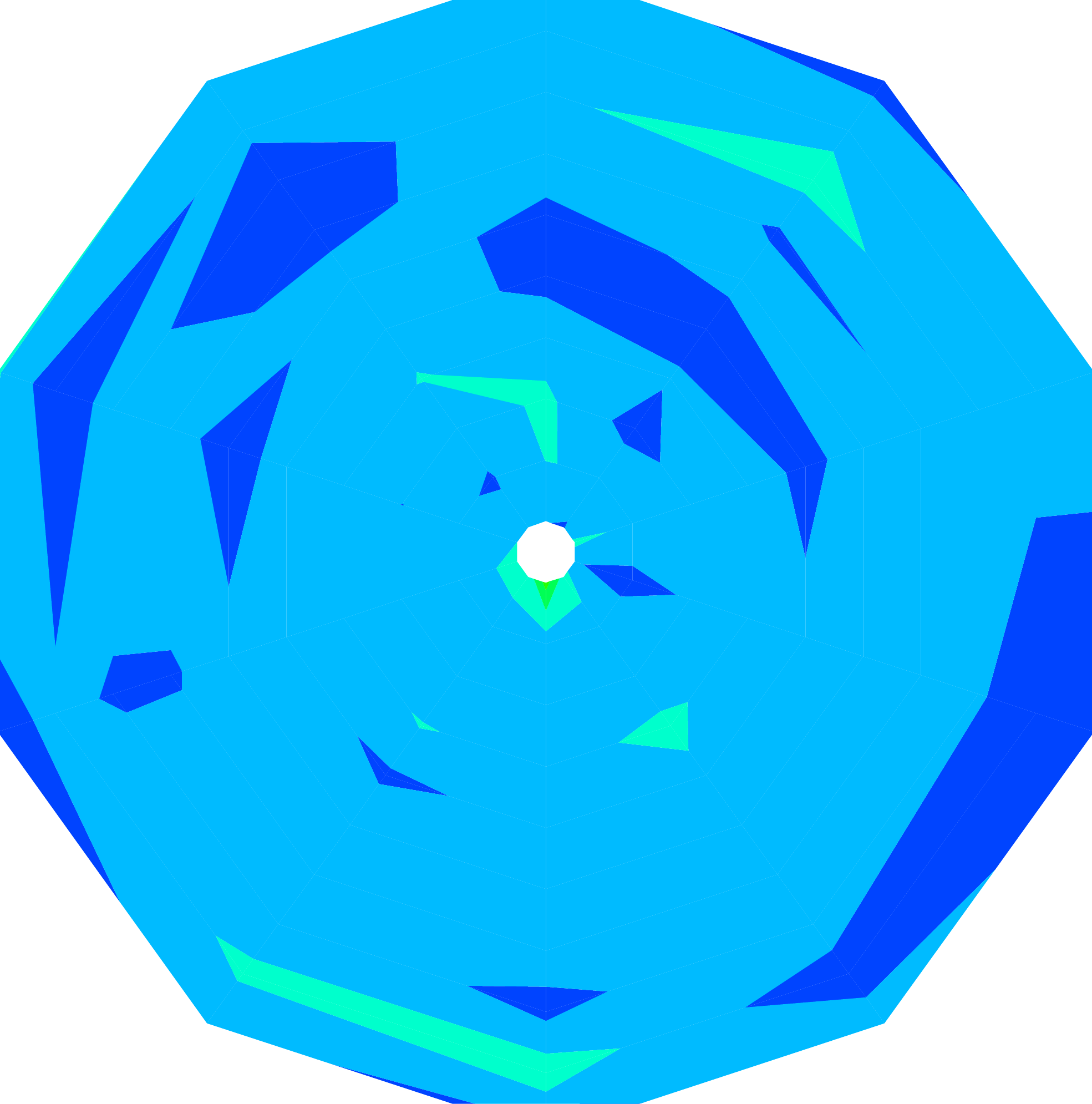}
            \caption{\footnotesize 1--9 days}
        \end{subfigure}
        \begin{subfigure}[b]{0.47\linewidth}
            \centering
            \includegraphics[width=\linewidth]{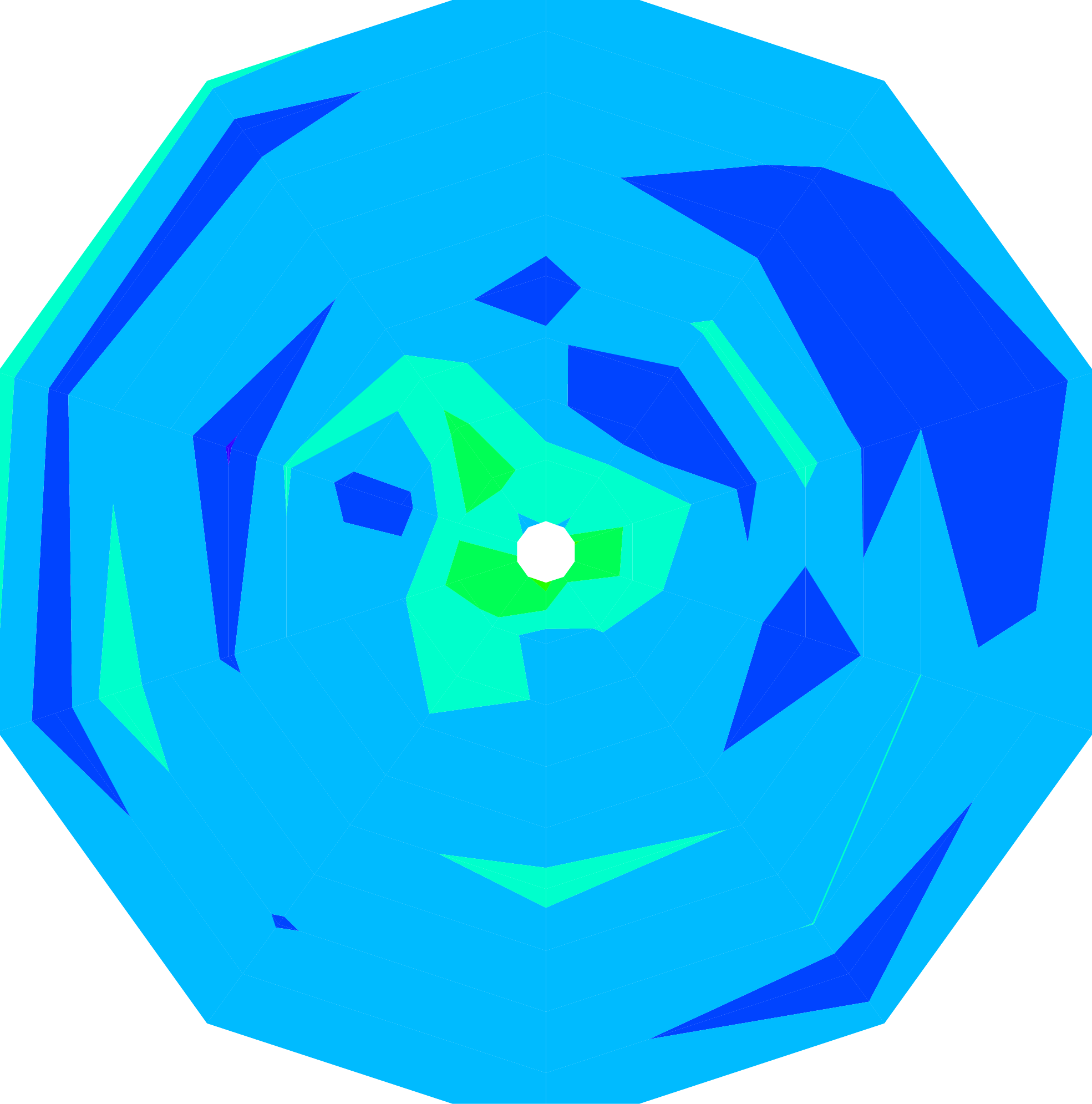}
            \caption{\footnotesize 9--25 days}            
       \end{subfigure}     
        \begin{subfigure}[b]{\linewidth}
             \color{white}
             \includegraphics[draft,height=0.05cm]{1_9Pre-injection_Scenarioangle}
             \color{black}   
        \end{subfigure}
        \begin{subfigure}[b]{0.47\linewidth}
            \centering
            \includegraphics[width=\linewidth]{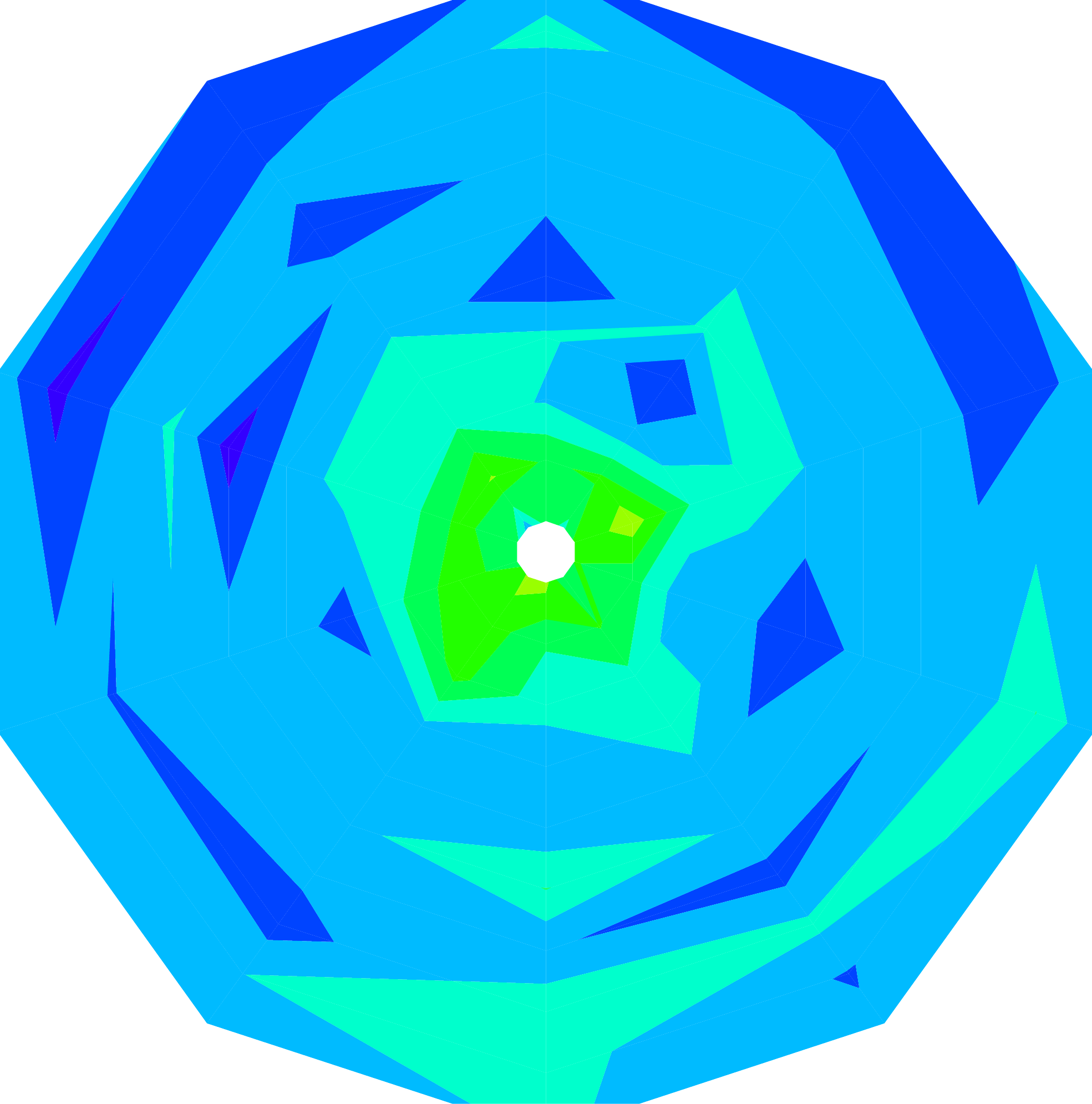}
            \caption{\footnotesize 25--49 days}            
        \end{subfigure}
        \begin{subfigure}[b]{0.47\linewidth}
            \centering
            \includegraphics[width=\linewidth]{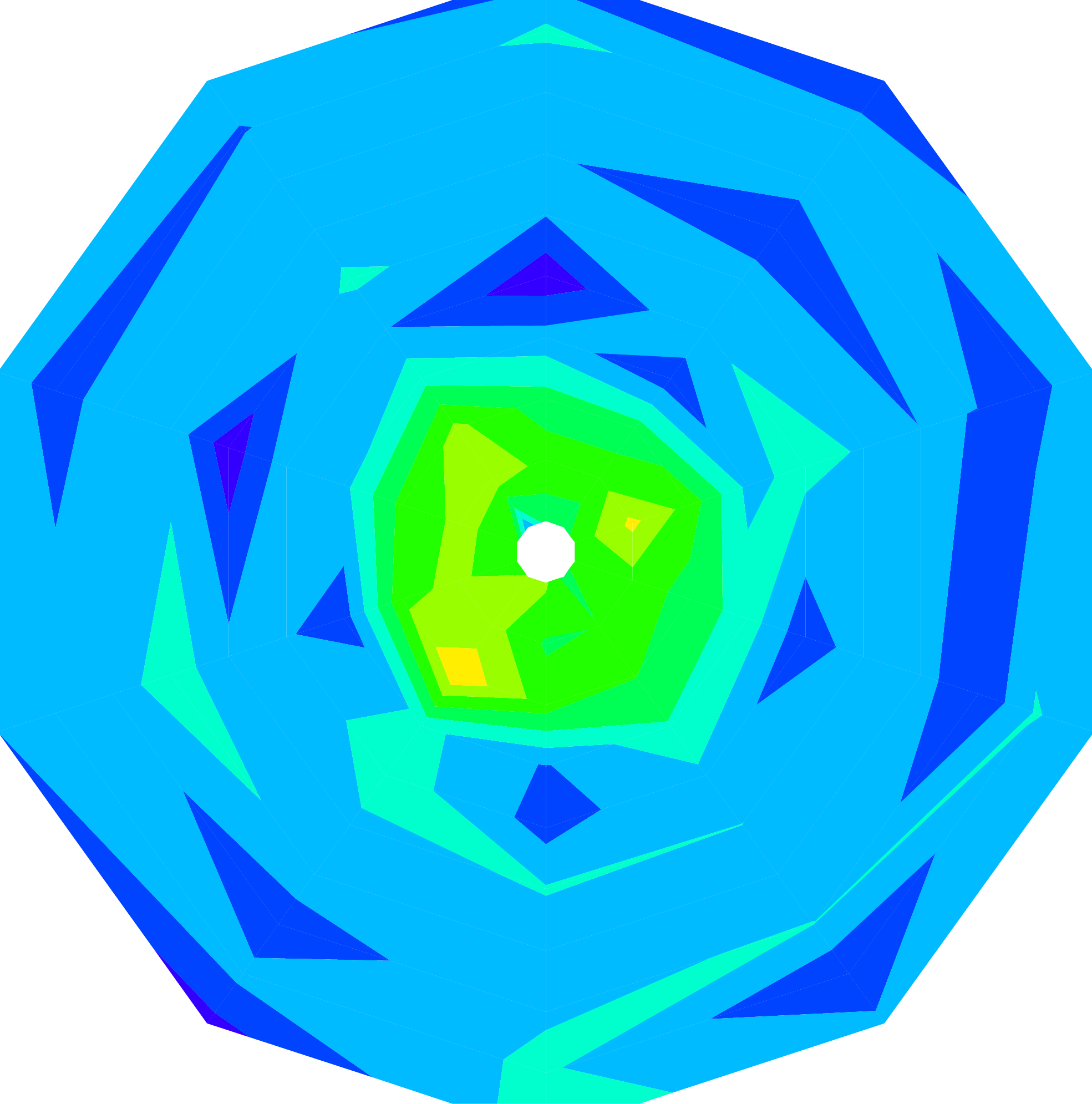}
            \caption{\footnotesize 49--81 days}            
        \end{subfigure}
        \begin{subfigure}[b]{\linewidth}
             \color{white}
             \includegraphics[draft,height=0.05cm]{1_9Pre-injection_Scenarioangle}
             \color{black}   
        \end{subfigure}             
        \begin{subfigure}[b]{0.47\linewidth}
            \centering
            \includegraphics[width=\linewidth]{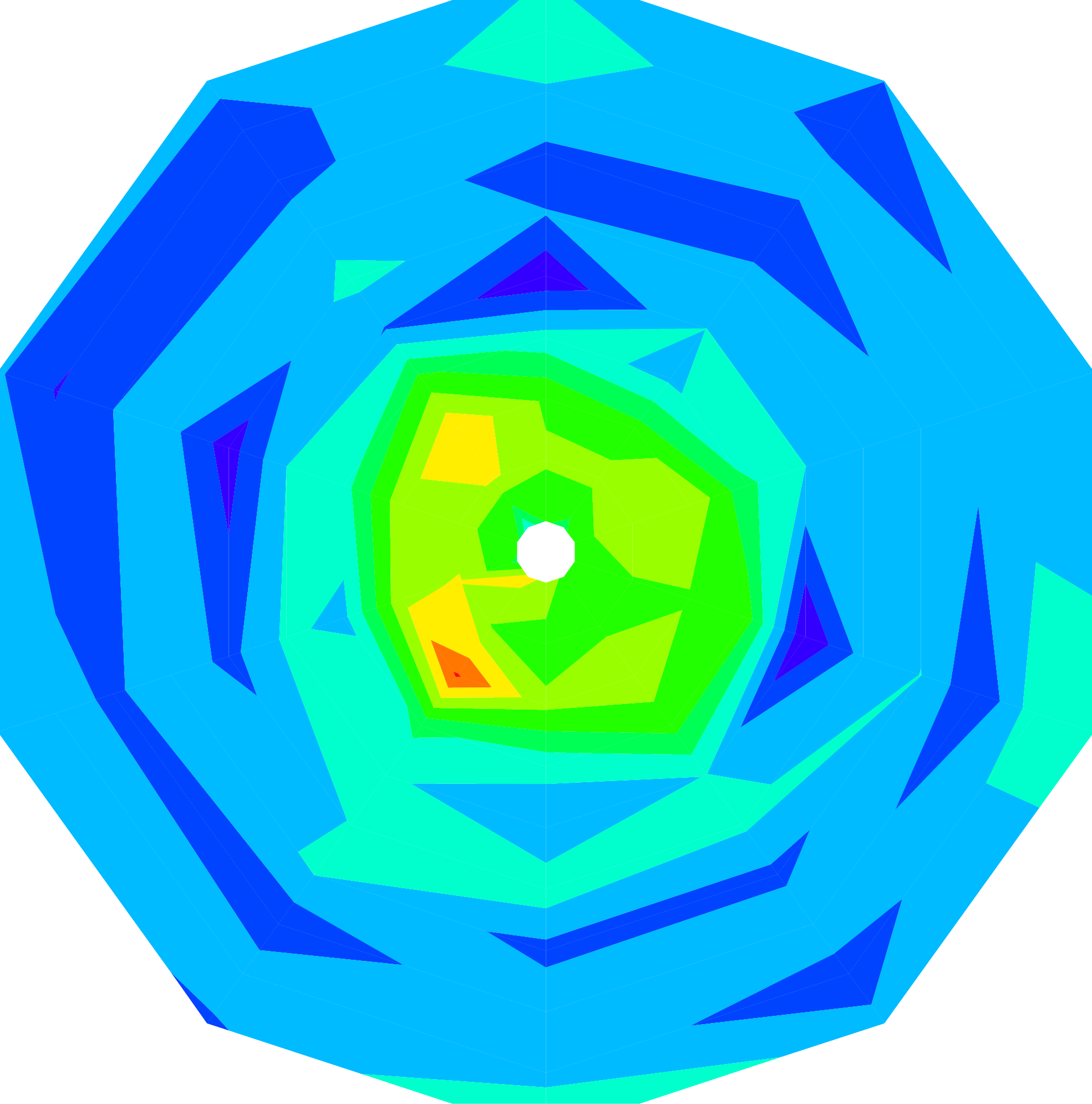}
            \caption{\footnotesize 81--121 days}            
        \end{subfigure}        
        \begin{subfigure}[b]{0.47\linewidth}
            \centering
            \includegraphics[width=\linewidth]{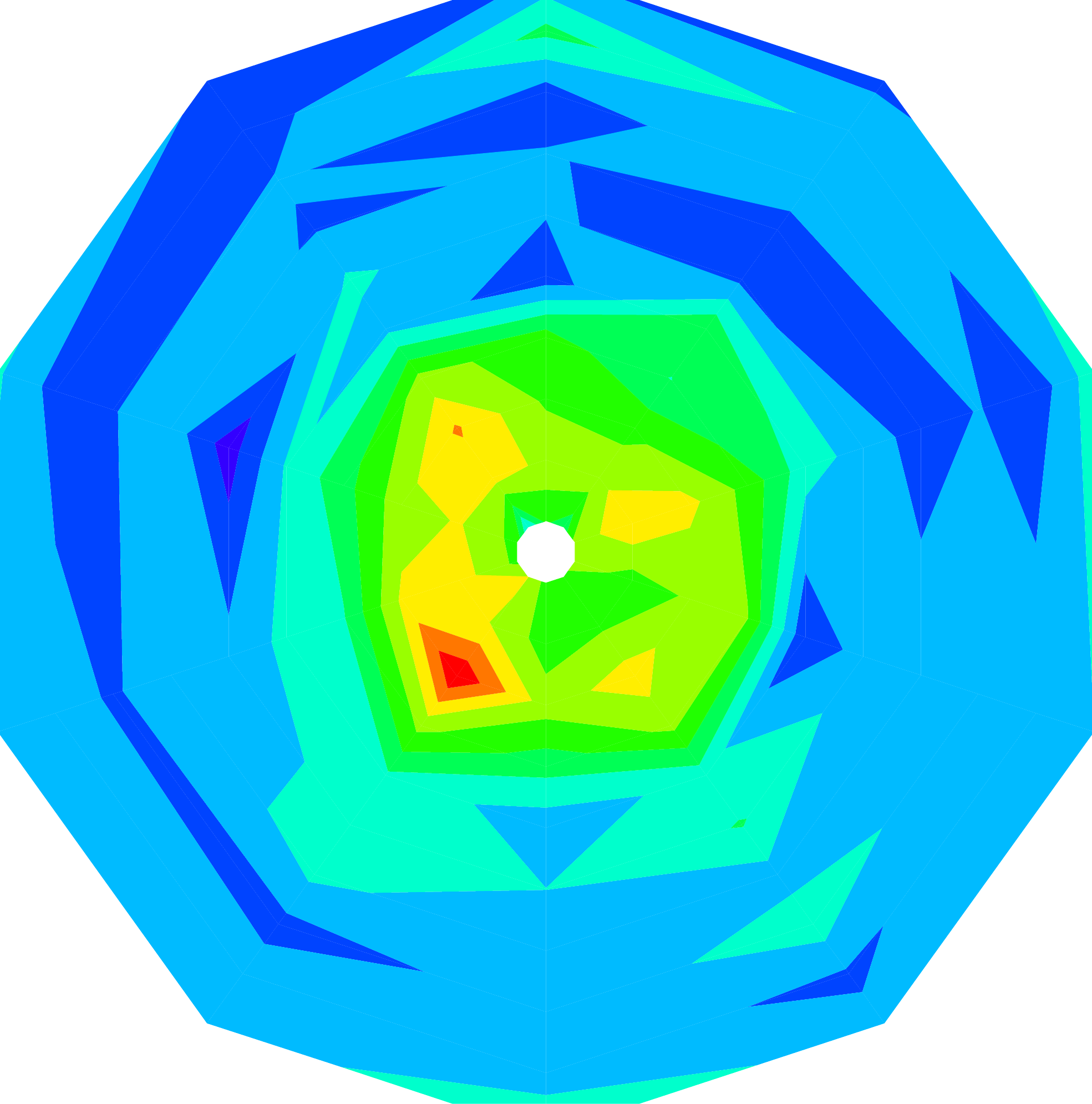}
            \caption{\footnotesize 121--169 days}            
        \end{subfigure}                         
 \end{subfigure}
     \begin{subfigure}[t]{0.2\linewidth}
           \begin{subfigure}[t]{\linewidth}
              \centering
              \includegraphics[width=\linewidth]{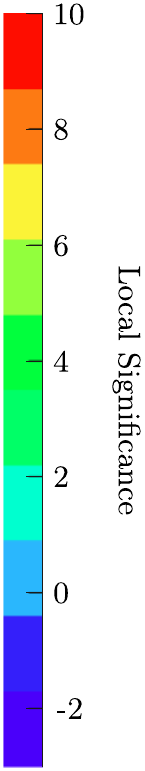}
           \end{subfigure}             
    \end{subfigure}    
\caption{The significance $S$ of the change in muon count in the detector plane 
due to the constant injection of \CO, using the model described in Section~\ref{plume}, 
as a function of muon zenith and azimuthal angle $\theta$ and $\psi$. The caption of each figure indicates the
time interval after first \co~injection. 
Each distribution contains 100 values of $S$, calculated in bins corresponding to $\Delta\theta = 7\degree$ radially and $\Delta\psi = 36\degree$ about the radial axis.
A linear interpolation is used between adjacent bins.}
\label{fig:Plume1}
\end{figure}

\begin{figure}[t]
    \begin{subfigure}[b]{0.75\linewidth}        
        \begin{subfigure}[b]{0.49\linewidth}
            \centering
            \includegraphics[width=\linewidth]{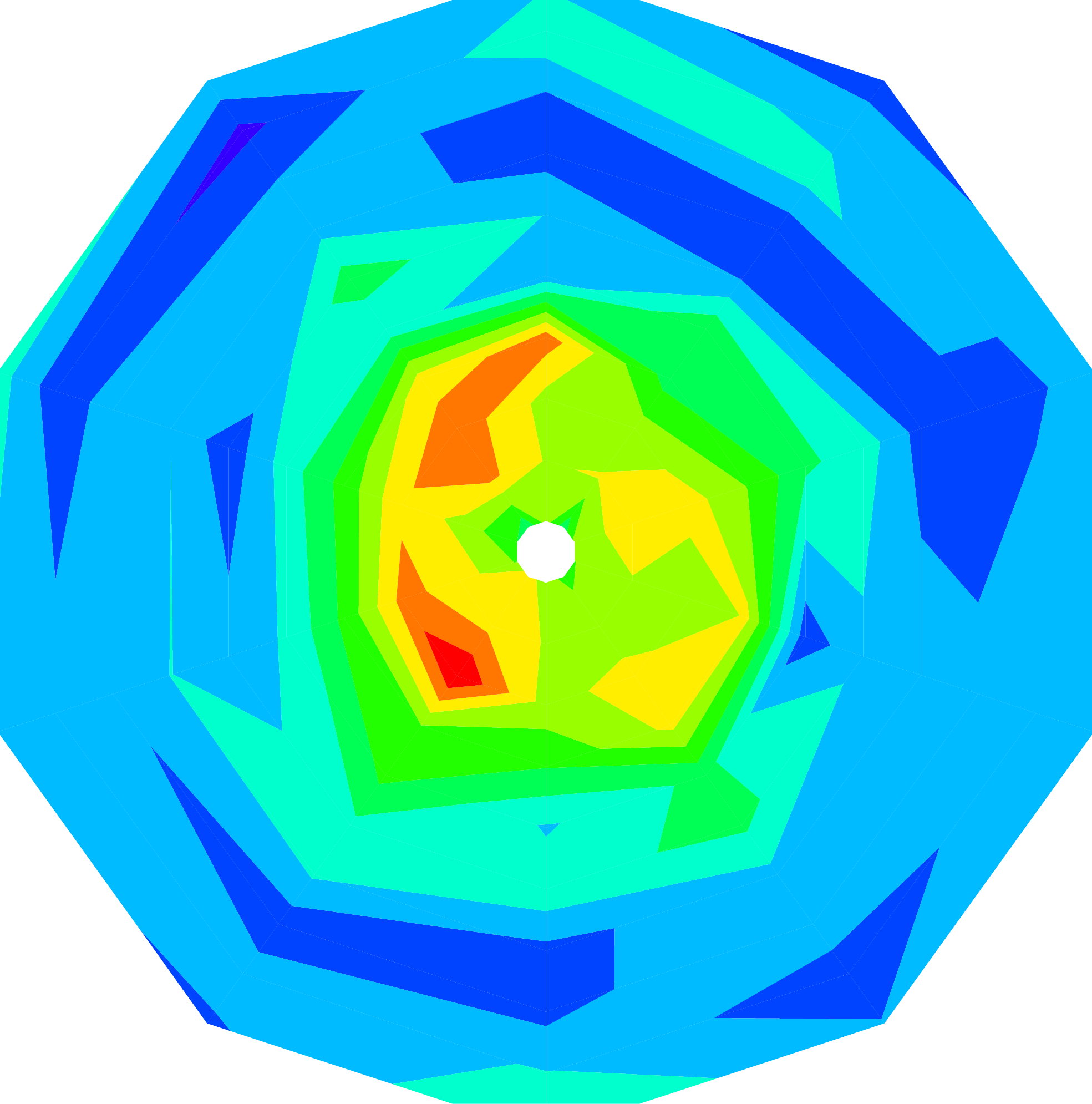}
            \caption{\footnotesize 169--225 days}            
        \end{subfigure}
        \begin{subfigure}[b]{0.49\linewidth}
            \centering
            \includegraphics[width=\linewidth]{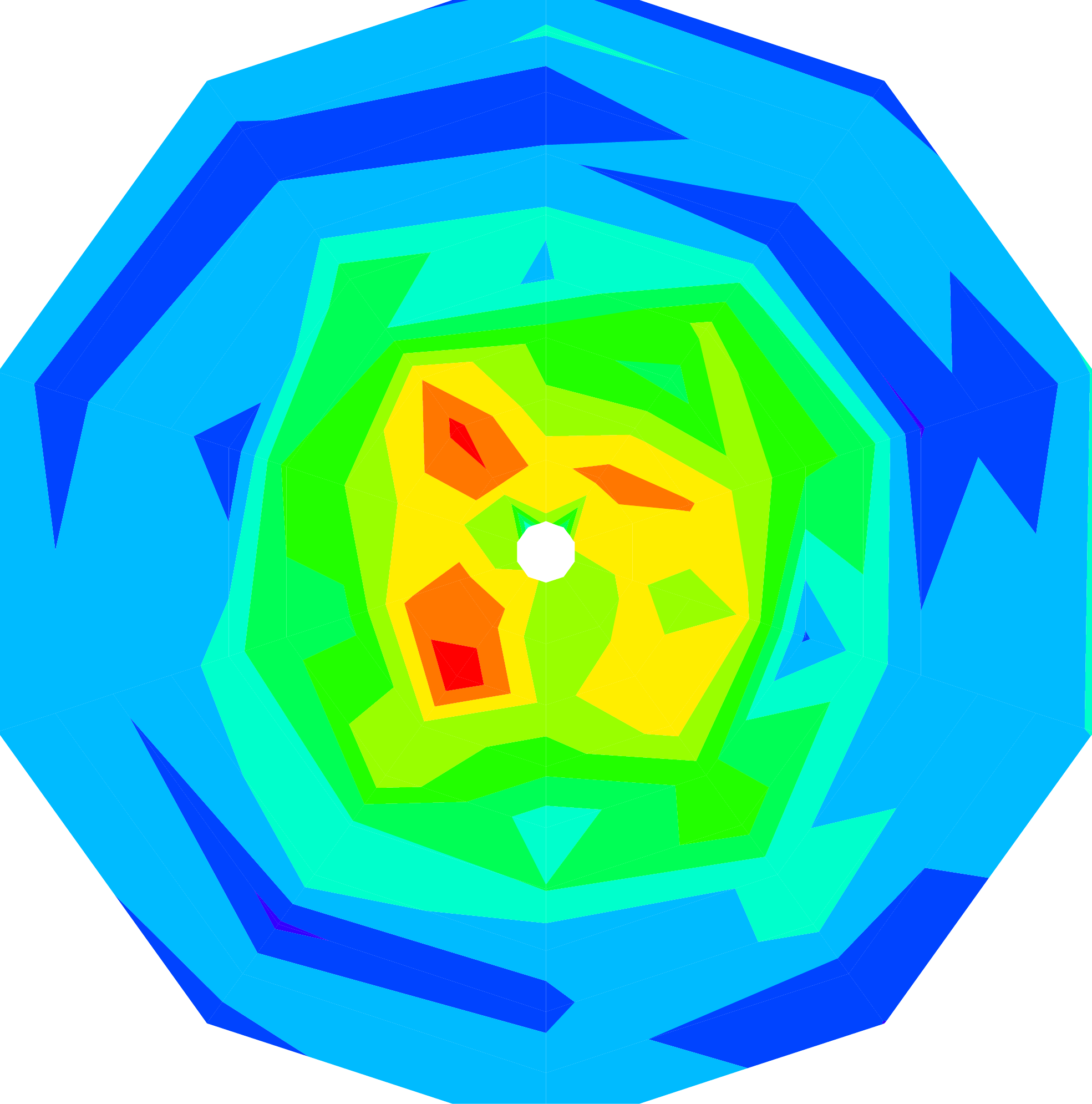}
            \caption{\footnotesize 225--289 days}            
        \end{subfigure}  
        \begin{subfigure}[b]{\linewidth}
             \color{white}
             \includegraphics[draft,height=0.1cm]{1_9Pre-injection_Scenarioangle}
             \color{black}   
        \end{subfigure}
        \begin{subfigure}[b]{0.49\linewidth}
            \centering
            \includegraphics[width=\linewidth]{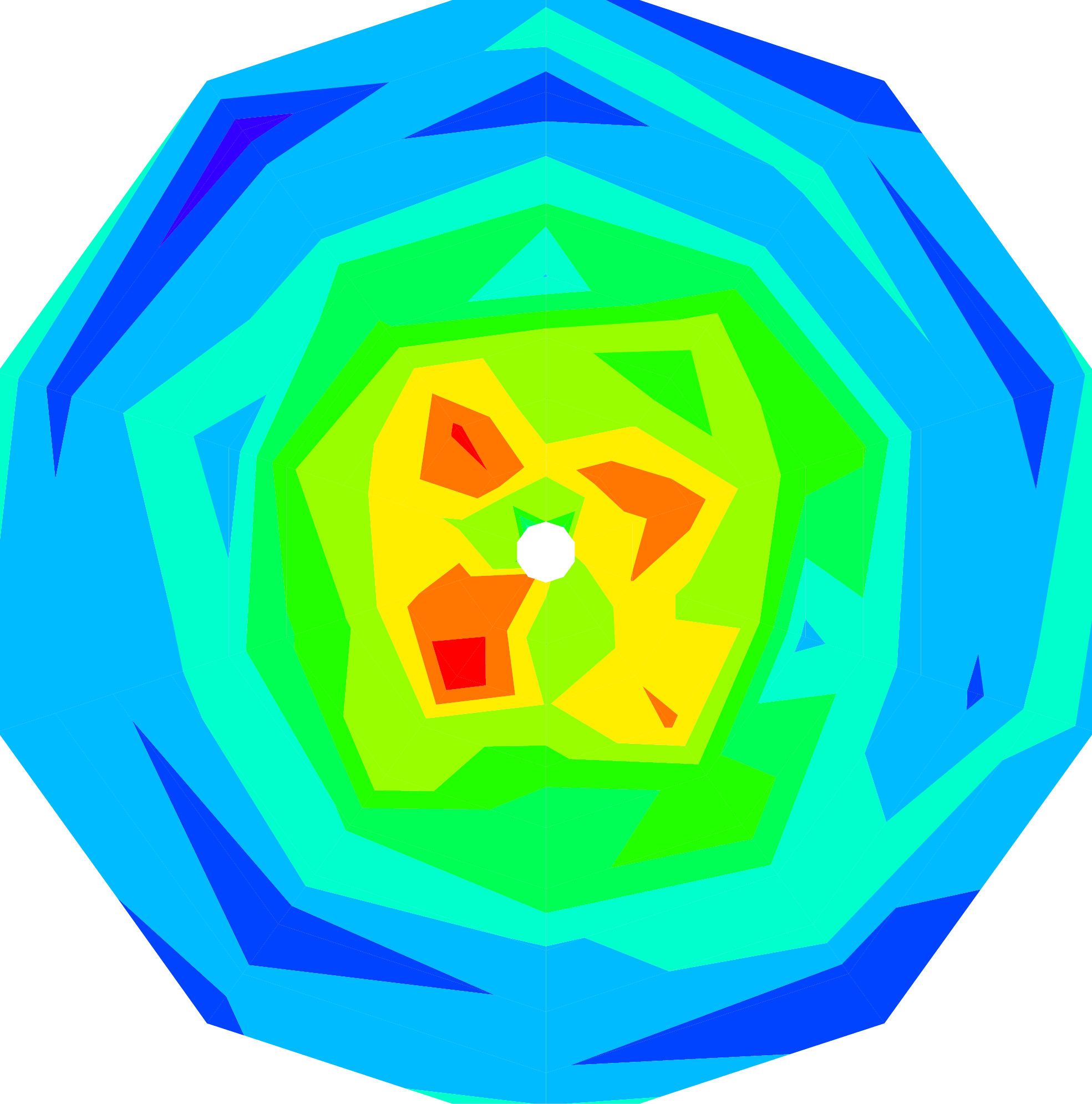}
            \caption{\footnotesize 289--361 days}            
        \end{subfigure}
        \begin{subfigure}[b]{0.49\linewidth}
            \centering
            \includegraphics[width=\linewidth]{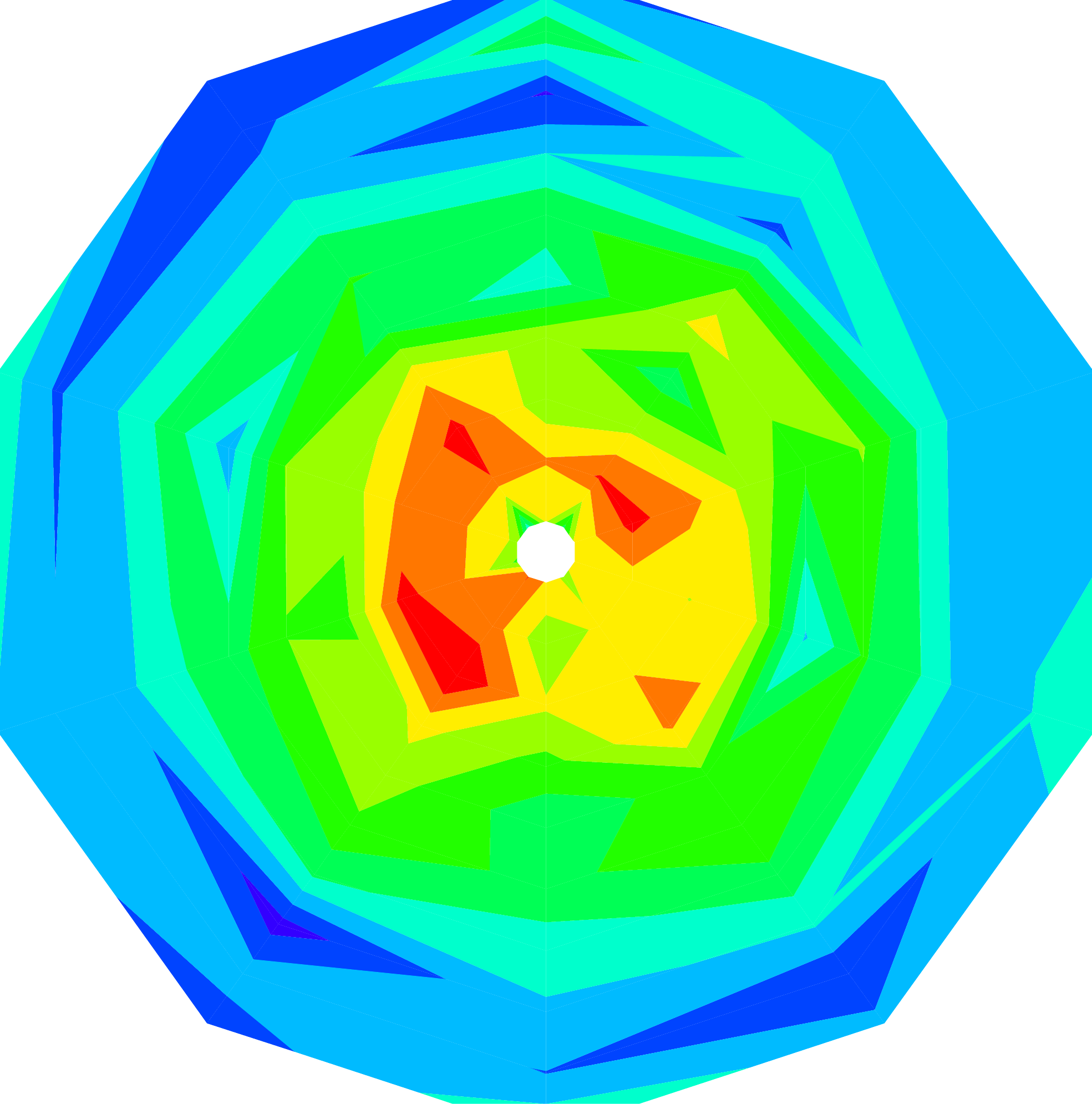}
            \caption{\footnotesize 361--441 days}            
        \end{subfigure}        
 \end{subfigure}
 \begin{subfigure}[b]{0.2\linewidth}
       \includegraphics[width=\linewidth]{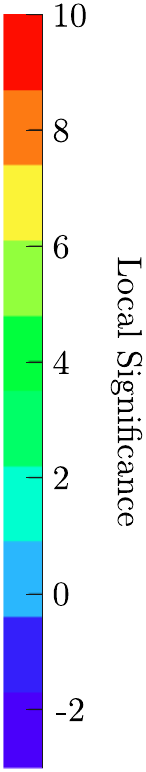}
 \end{subfigure}
\caption{The significance $S$ of the change in muon count in the detector plane 
due to the constant injection of \CO, using the model described in Section~\ref{plume}, 
as a function of muon zenith and azimuthal angle $\theta$ and $\psi$. The caption of each figure indicates the
time interval after first \co~injection. 
Each distribution contains 100 values of $S$, calculated in bins corresponding to $\Delta\theta = 7\degree$ radially and $\Delta\psi = 36\degree$ about the radial axis.
A linear interpolation is used between adjacent bins.}
\label{fig:Plume2}
\end{figure}

\begin{figure}[h]
	\begin{subfigure}[b]{\linewidth}
		\includegraphics[width=\linewidth]{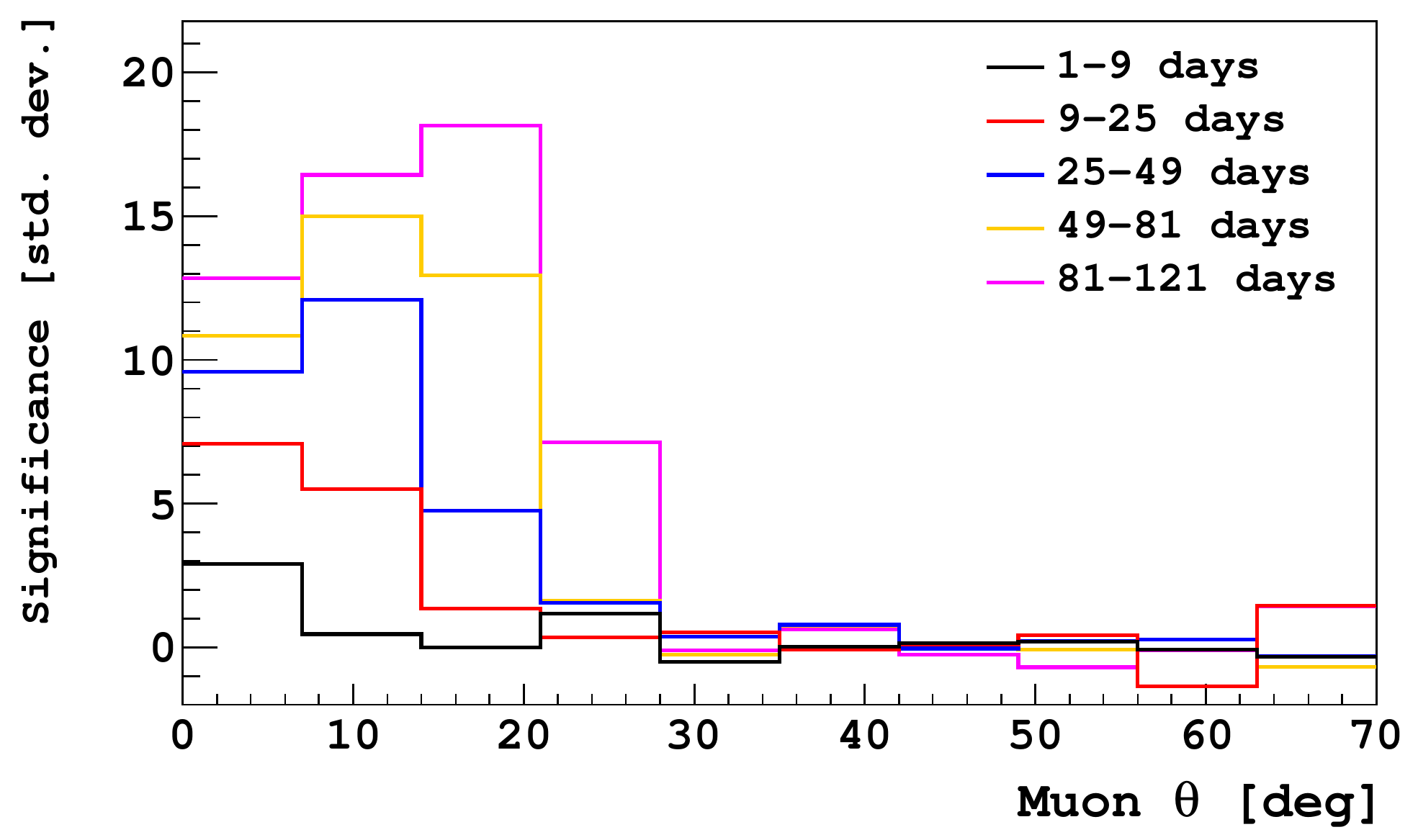}
	        \caption{\footnotesize 1 -- 121 days}
	\end{subfigure}
	\begin{subfigure}[b]{\linewidth}
		\includegraphics[width=\linewidth]{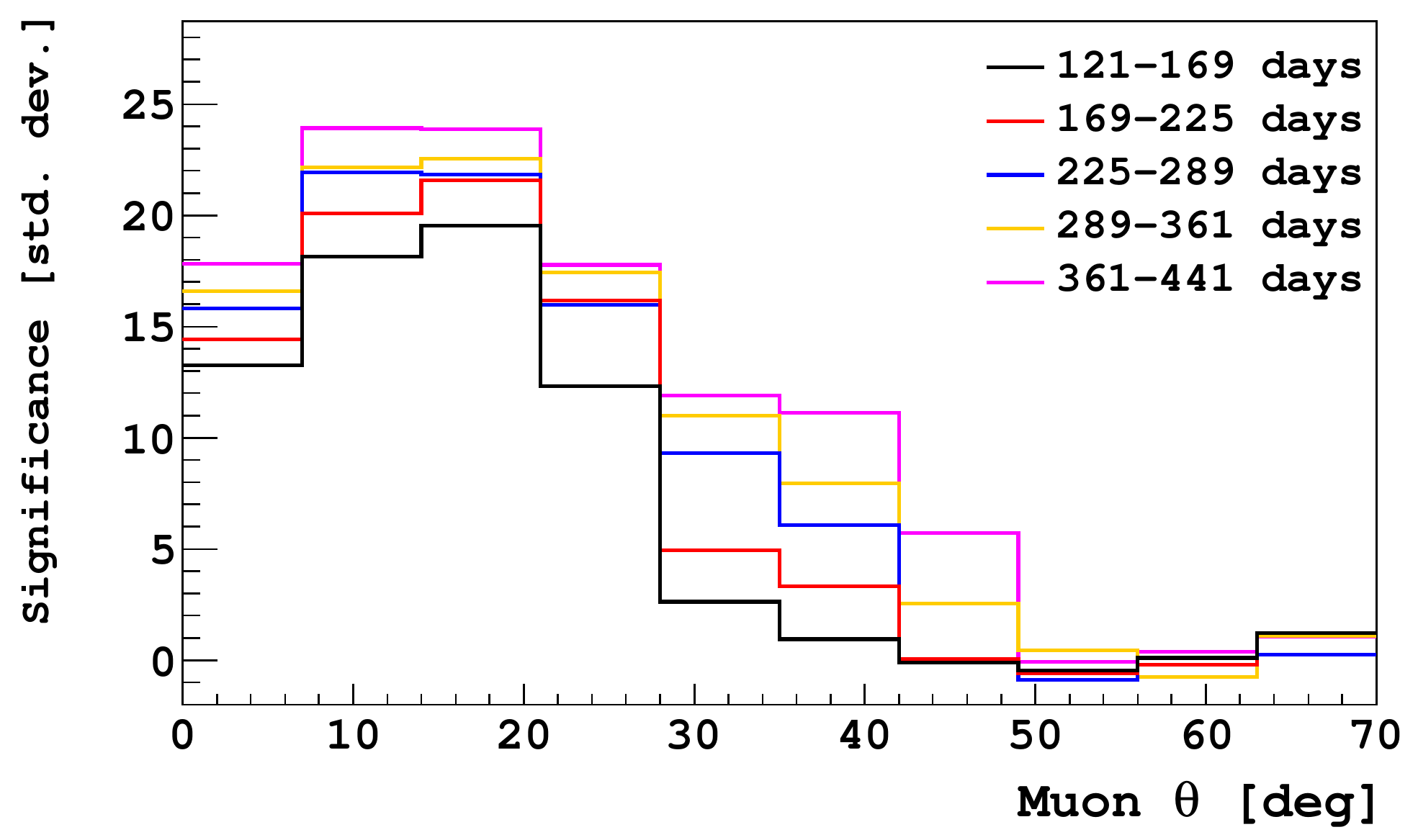}
	        \caption{\footnotesize 121 -- 441 days}
	\end{subfigure}
	\caption{The significance $S$ of the change in muon count in the detector plane 
due to the constant injection of \CO, using the model described in Section~\ref{plume}, 
as a function of muon zenith angle $\theta$. The caption of each of the two figure indicates the
time interval after first \co~injection.}
\label{plumetheta}
\end{figure}

The angular distribution of $S$
is shown in Figures~\ref{fig:Plume1} and~\ref{fig:Plume2} for the periods 0--169~days and 169--441 days after first \co~injection, respectively. 
The shape of the plume distributions is visible above the background of statistical fluctuations, even after 25 days. 
Figure~\ref{plumetheta} shows the distribution of $S$ as a function of $\theta$ only. The value of $\theta$ corresponding to the maximum value of $S$ is
shown to shift as a function of time, which is related to the dynamics of the \co~plume and also the sensitivity to the muon path length.

\begin{table}[h]
\footnotesize
\renewcommand{\arraystretch}{1.2} 
\caption{The muon flux $\Phi_\mu$ as a function of time step $T$ (or equivalently the radial extent of the \co~plume, 
$R$) and the associated significance $S$ relative to the muon flux prior to injection. The presented uncertainties are taken from the statistical uncertainty in the muon count.}
\centering
\begin{tabularx}{\linewidth}{c *{4}{Y}}
\toprule
$T$ [days] & $R$~[arb.~units] & $\Phi_{\mu}$~[$10^{-7}$cm$^{-2}$s$^{-1}$] & $S$
\\\midrule 
1--9 & 1 &  $ 2.3185~\pm~0.0018 $ & 0.98 \\
9--25 & 2  & $ 2.3236~\pm~0.0013 $ & 3.9 \\
25--49 & 3  & $ 2.3278~\pm~0.0010 $ & 24 \\
49--81 & 4 & $ 2.3302~\pm~0.0009 $ & 32 \\
81--121 & 5  & $ 2.3348~\pm~0.0008 $ & 40 \\
121--169 & 6 & $ 2.3382~\pm~0.0007 $ & 48 \\
169--225 & 7  & $ 2.3410~\pm~0.0006 $ & 56 \\
225--289 & 8  & $ 2.3426~\pm~0.0006 $ & 64 \\
289--361 & 9  & $ 2.3443~\pm~0.0006 $ & 72 \\
361--441 & 10  & $ 2.3461~\pm~0.0006 $ & 80
\\\bottomrule
\end{tabularx} 
\label{tab:plume_tab}
\end{table}

\section{Discussion}
\label{disc}

In order to perform this simulation, we have chosen parameters which are applicable to the geology in the vicinity of Boulby Mine, Cleveland, United Kingdom. 
In realistic off-shore sites several parameters will change, and different analysis strategies may be adopted.
This section presents a short discussion in order to provide a platform for future studies. Furthermore, we present a brief discussion on the practical viability
of this technology, in particular with respect to the relative costing compared to existing technologies.


For this study we have chosen a conservative value for the mass-rate of \co~injection, of 20 kg/s ($\sim 0.63$~Mtpa). Therefore it is clear that
for realistic implementations of this technology, the sensitivity to both the plume extent and density variations will increase.

Clearly the depth of the \co~storage reservoir will vary between sites, and therefore
the depth of the muon detector array will also vary between sites. 
Realistic off-shore sites, such as the Hewett Fields complex in the North Sea, have reservoir depths $\gtrsim 1.2$~km.
Compared to the depth used in this simulation ($\sim 0.8$~km) this corresponds to a reduction in the muon flux of 85\%, based on the study presented in Ref.~\cite{Kudryavtsev2009339}, which therefore reduces $S$ by a factor of $\sim 0.39$.
This does not significantly alter the conclusions of this study.

After significantly longer storage times the lateral plume size will extend well beyond what is discussed in this paper. 
Experience from the Sleipner CCS project~\cite{Eiken20115541},
which operates in an offshore deep saline aquifer (depth $\sim0.8$~km), suggests that \co~plumes may have a lateral range of approximately 15~km after 14~years. 
In order
to image a plume with such a lateral range, 
one needs several arrays of muon detectors positioned at different sites relative to the injection point. 
Several arrays of detectors are also required to reconstruct a proper tomographic image of the \co~migration (in addition to the `line-of-sightÕ reconstruction). 
This will allow for a triangulation of the muon
data, in order to reconstruct a spatially three-dimensional image of the \co~plume. With the detection strategy that is presented in this paper,
which only uses a single detector array, one may still interpret vertical movements from unexpected changes in the angular distribution. 
The manner in which this process is optimised will vary from site to site.

One limiting factor, in terms of the sensitivity of this technique, is the choice of packing fraction \dpack~of the borehole containers 
across the total surface area of the detector array. Realistic values for this parameter are currently unclear,
although this may not be important if a smaller packing fraction can be compensated by a larger area for detector deployment.

As the effectiveness of muon radiography for the mapping of large scale objects is already proven (Ref.~\cite{alvarez1,Marteau201223,1995998}),
the utility of this technology will ultimately depend upon the practicality of deployment of the technology at \co~injection sites and 
the cost of deployment relative to other monitoring technology which could be used. Clearly the practicality of deployment will require further investigation. 
We can however provide a rough estimate of the relative cost of muon radiography, with respect to
seismic technology, based on a review of freely available information and without reference to particular vendors.
We choose to compare to seismic surveying, as it is clearly the most widely used monitoring technology for off-shore injection sites.
The relative evaluations are based on a 2015 cost base. 

Before we present our cost evaluation, we emphasise a numbers of points:

\begin{itemize}
\item Costs in this industry are volatile although relative costs, as we will discuss, are approximately fixed. Precise cost estimates are impossible on this basis.
\item Multilateral sidetrack technology, which is discussed in this section, may be used for deploying the detectors, 
although further research may yield more favourable or cost-effective methods.
\item In the early stages of this technology we suggest to use modified injection wells, rather than drilling new wells for monitoring.
\item Muon detector technology is expected to drop significantly in price if large-scale deployment is implemented, although the scale of this reduction is difficult to quantify.
\end{itemize}

In the few sites which have been developed for off-shore \co~storage~\cite{Eiken20115541}, 
seismic surveys have been performed ahead of any injection and subsequently on an annual basis. 
The cost of such surveys varies widely and is responsive to the oil price, 
since most seismic data will be acquired to enable mapping of petroleum accumulations and their response to production.  
A typical single, modestly sized survey of around 100~km$^2$ will today cost in the region of \$5 million. 
For a storage site lasting 40 years this equates to \$200 million over the duration of the injection phase. This will 
increase further if one considers post-injection monitoring, processing and reprocessing of the annual data.

The cost to deploy muon sensors is similarly difficult to assess due to the same volatility issues on oil price. Using the same cost base we can estimate the incremental cost for 
deepening a well and deploying muon detectors beneath the injection site. A mid-price semi-submersible or jack-up rig for the North Sea will today cost in the region of \$300,000 per 
day. For the initial deployment of detectors the mobilisation and demobilisation costs will fall to the injection well itself and can be discounted for these comparison purposes. 
Detectors would be deployed in multilateral sidetracks drilled from the mother borehole beneath the injection site, in a process known as coiled tubing drilling~\cite[see, for instance,][and references therein]{Afghoul}.
Coiled tubing drilling is used extensively in off-shore settings to increase the drainage surface of well-bores into an oil reservoir.  A mother borehole 
may have several tens of daughter boreholes.
For the purpose of muon radiography, it is likely that short radii sidetracks would be drilled. We could consider 18 sidetracks 
(which allows for a total instrumented surface area of 1000~m$^2$)
which are each constructed over two days, plus two weeks to deploy and complete the instrumented part of the well array. 
This equates to about 50 days at \$300,000 or \$15~million in total. 

One must then consider the cost for developing a set of detectors that would instrument the required 1000~m$^2$ area. 
We assume that a single detector occupies a surface area of 0.2~m$^2$, and contains
approximately 25 scintillator bars, each 1~m long. With current (2015) market costs within a laboratory, one could
acquire all scintillator bars (in one detector) for \$4,000, photo-sensors for \$6,000, a specially designed data acquisition system for
\$2,000, other components for \$1,000 and a further \$7,000 for technical labour costs. One could envisage a cost reduction 
factor of at least two due to mass production - which would imply a cost of about \$10,000 per detector. Therefore in order to 
instrument the required 1000~m$^2$ area, this implies a cost of \$50 million. Again, the uncertainties in the precise costing at this stage are clearly very large so 
this figure may rise or fall, but an estimate of the order of tens of millions of dollars is applicable. 
We are confident that significant cost savings will be possible by using novel and cheaper alternatives to various detector components, 
and a further cost reduction by mass production of detectors, when this technology becomes widely used on an industrial scale for geological repositories and CCS.

Based on experience in experimental particle physics experiments, one can expect scintillator detectors to 
operate for at least ten years, necessitating approximately four sets of detectors over the 40 year injection period. 
In practice though, seismic would still be deployed to detect plume extent 
in the event that the plume can be seen to be moving using muon radiography. One should note that the cost of both seismic acquisition and drilling are highly volatile and both are currently falling in line with a low oil price, however the ratio between drilling and seismic costs is effectively stable. 

\section{Conclusion}

We have shown the results of the first detailed simulation of muon radiography for the purpose of the monitoring of
\CO~stored in geological CCS repositories. 
Our model is the first to incorporate 
geological data to describe a \CO~storage site, with a simplified but nevertheless realistic modelling of \co~plume evolution and muon detector configuration.
Non-ideal behaviour in real situations is to be expected because of geological heterogeneity or other factors, for example, variable wettability of mineral surfaces. Simulations like this will become an integral part of any future use of this CCS technology, to compare actual outcomes with the initial reservoir model.

We use a detailed numerical model of the fluid flow of \CO~to show that muon radiography is sensitive to 
the time-evolution of a \co~plume formation. We have shown that after approximately 49~days of constant \co~injection, 
muon radiography has a strong sensitivity to changes in column density of $\lesssim 1\%$, in the scenario that we have presented.

From our studies, it is clear that this technology is a conceptually viable monitoring method, 
and in combination with the study of a full prototype detector, 
represents a crucial milestone in the realisation of muon radiography in the context of CCS.

\section{Acknowledgements}

This work was supported by the Department of Energy and Climate Change and Premier Oil plc.
We would like to thank the Science and Technology Facilities Council (STFC, UK) for supporting this work (grant ST/K001841/1). The studentship of D.~Woodward is funded by STFC (grant ST/L502492/1). 
The contribution of M.~Coleman was carried out in part at the Jet Propulsion Laboratory (JPL), California Institute of Technology, under contract with the National Aeronautics and Space Administration (NASA).
We would like to acknowledge the continued support of Israel Chemicals Ltd UK at Boulby Mine and the team
at the STFC Boulby Underground Laboratory.  We thank ENI for providing data relating to the Hewett Fields complex in the North Sea. 
We also thank David Jacques and Tom Lynch, from the University of Leeds, for useful discussions.

\FloatBarrier

\bibliography{Muon_CCS_Simulations.tex}

\end{document}